\documentclass[a4paper,fleqn,usenatbib]{mnras}
\hypersetup{draft}

\usepackage{newtxtext,newtxmath}

\usepackage[T1]{fontenc}
\usepackage{ae,aecompl}

\usepackage{graphicx}	
\usepackage{amsmath}	
\usepackage{lineno,hyperref}
\usepackage{pdflscape}
\usepackage{multirow}
\usepackage{wasysym}
\usepackage{longtable}

\usepackage{siunitx}
\usepackage[T1]{fontenc}
\usepackage[utf8]{inputenc}
\usepackage{gensymb}
\usepackage{wasysym}

\DeclareMathOperator{\arctantwo}{arctan2}

\usepackage[colorinlistoftodos]{todonotes}
\let\Oldtodo\todo
\renewcommand{\todo}[1]{\Oldtodo[inline]{#1}}

\title[Optical Fluxes]{Computing optical meteor flux using Global Meteor Network data}

\author[D. Vida et al.]{
Denis Vida,$^{1,2}$\thanks{E-mail: dvida@uwo.ca}
Rhiannon C. Blaauw Erskine$^{3}$,
Peter G. Brown$^{1,2}$,
Jonathon Kambulow$^{1,4}$,
\newauthor
Margaret Campbell-Brown$^{1, 2}$,
Michael J. Mazur$^{1}$
\\
$^{1}$Department of Physics and Astronomy, University of Western Ontario, London, Ontario, N6A 3K7, Canada\\
$^{2}$Institute for Earth and Space Exploration, University of Western Ontario, London, Ontario, N6A 5B8, Canada\\
$^{3}$Jacobs Space Exploration Group, Marshall Space Flight Center, Huntsville, Alabama, 35812, USA \\
$^{4}$Department of Physics and Astronomy, University of Waterloo, Waterloo, Ontario, N2L 3G1, Canada \\
}

\date{Accepted 2022 June 21. Received 2022 June 21; in original form 2022 June 05.}

\pubyear{2022}

\begin{document}
\label{firstpage}
\pagerange{\pageref{firstpage}--\pageref{lastpage}}
\maketitle

\begin{abstract}
Meteor showers and their outbursts are the dominant source of meteoroid impact risk to spacecraft on short time scales. Meteor shower prediction models depend on historical observations to produce accurate forecasts. However, the current lack of quality and persistent world-wide monitoring at optical meteoroid sizes has left some recent major outbursts poorly observed.
A novel method of computing meteor shower flux is developed and applied to Global Meteor Network data. The method is verified against previously published observations of the Perseids and the Geminids. The complete mathematical and algorithmic details of computing meteor shower fluxes from video observations are described.
As an example application of our approach, the flux measurements of the 2021 Perseid outburst, the 2020-2022 Quadrantids, and 2020-2021 Geminids are presented. The flux of the 2021 Perseids reached similar levels to the 1991-1994 and 2016 outbursts (ZHR $\sim$ 280). The flux of the Quadrantids shows high year-to-year variability in the core of the stream while the  longer lasting background activity is less variable, consistent with an age difference between the two components. The Geminids show a double peak in flux near the time of peak.

\end{abstract}




\section{Introduction} \label{sec:introduction}

Modelling and measuring the flux and associated risk of meteoroids to spacecraft in and beyond Earth’s orbit is essential to all space operations, especially human spaceflight programs \citep{moorhead2020realistic, moorhead2021meteor}. Space agencies produce meteor shower forecasts \citep{moorhead2020nasa}, which predict the number of meteoroids above a given mass threshold that may impact a spacecraft over a particular time period.

The sporadic complex is the dominant impact risk to spacecraft over long time intervals \citep{moorhead2019meteor}, however the shower meteoroid flux may dominate the impact risk compared to sporadics in some short periods (hours to days). Moreover, as shower velocities are typically much higher than those of sporadic meteoroids, the threshold shower mass for a given kinetic damage is smaller while the risk from impact plasma is much higher \citep{moorhead2019meteor}. 

The three strongest annual showers at Earth are the Perseids, Geminids, and Quadrantids, each regularly displaying a Zenithal Hourly Rate (ZHR) of over 100 \citep{moorhead2019meteor}. The Perseids and Geminids are comparatively older ($>$1000s of years) streams which have been studied widely \citep{brown1998simulation, blaauw2016optical, de2010origin}. Apart from rare Perseid outbursts, they display predictable annual behavior \citep{miskotte2017magnificent, jenniskens2021perseid}. The Quadrantids, however, are a young, short-duration shower known to display large year-to-year variations \citep{belkovich1974determination, rendtel2016quadrantids, abedin2015, abedin2018} making it the least well documented of the major showers. 

The Perseids, whose parent body is the comet 109P/Swift–Tuttle, is arguably the most well known meteor shower. It consists of an old ($\sim25,000$ yrs) core component which produces regular returns with a ZHR of $\sim100$ \citep{rendtel2020handbook}. In addition, a younger (100s yrs) component \citep{brown1998simulation} produces infrequent outbursts with ZHRs as high as 300 \citep{brown1996perseid}. The outbursts are caused by direct gravitational perturbations from Jupiter and Saturn which shift the nodes of Perseid meteoroids inward, making them collide with Earth \citep{brown1998simulation}. The outbursts significantly increase  meteoroid impact risk to spacecraft \citep{beech1993impact} because the meteoroids are fast, and even $\sim10^{-5}$ g meteoroids can cause significant damage \citep{moorhead2019meteor}. A Perseid meteoroid from the 1993 outburst is considered the likely cause of the demise of an operational satellite \citep[Olympus-1;][]{caswell1995olympus}.

Unlike other meteor showers whose parent bodies are comets, the Geminids originate from the C-complex asteroid 3200 Phaethon. Their ablation behaviour in the atmosphere is also distinct and consistent with stronger asteroidal material \citep{ceplecha1998meteor}. A major outstanding question related to the shower is its mode of formation and the ultimate nature of Phaethon - is it an asteroid or a defunct comet \citep{Ryabova2021}? Key to unravelling this mystery is the structure of the Geminid stream, in particular its current activity profile. 

The mass of the stream is estimated to be as high as $\sim10\%$ of Phaethon's \citep{hughes1989mass, blaauw2017mass}, but large uncertainties remain \citep{ryabova2017mass}. The asteroid currently shows insufficient activity and mass loss to explain the mass residing in the stream \citep{jewitt2010activity, tabeshian2019asteroid}. However, to accurately estimate the stream mass, the structure of the stream and the way the Earth intersects it must be known. The exact shape of the activity profile directly informs these, which is why flux measurements are desirable \citep{ryabova2007mathematical}.

The Quadrantid meteor shower was first observed in 1835, but modelling suggests that the broader shower has been active for several thousand years. The core of the stream where the flux is highest is only two centuries old \citep{wiegert2005, abedin2015}. The proposed parent body of the core is asteroid 2003 EH1, though other objects have been linked to the Quadrantids, most notably 96P/Machholz and C/1490 Y1 \citep{abedin2015, abedin2018}. The shower displays a sharp flux peak around $283^{\circ}$ degrees solar longitude (corresponding to approximately January 3) with half width of only about 6-8 hours. Thus in the visual size ranges, the Quadrantids are only seen for one night, and mostly in the northern hemisphere which suffers from poor weather at this time of the year. 

Generally, absolute measurement of meteoroid flux has been made from in-situ impacts \citep{Grun2002} as well as radar \citep{Kaiser1960}, visual \citep{koschack1990determination1} and video \citep{molau2013meteoroid, molau2020meteorflux} meteor observations. Each technique has inherent biases and significant effort is required to develop robust flux measurements from raw values. In particular, though there has been explosive growth in the deployment of video meteor systems in the last decade \citep{koten2019meteors}, the video meteor flux estimation methodology has often been only briefly described in prior literature \citep[e.g.][]{Brown2002b, gural2002meteor, campbell2006annual, blaauw2016optical}.

Here we describe, validate and justify corrections and analysis procedures to be applied to single-station video meteor measurements to convert raw observed counts to meteor shower fluxes. The algorithm is generally applicable to any optical system, but here we focus on fluxes as measured by the Global Meteor Network (GMN). In our approach, single-station observations from multiple video systems are aggregated into an equivalent 'global' instrumental meteor detector to improve statistics and temporal resolution. We apply this procedure to the Perseids, Quadrantids, and Geminids to compare with other measurements and verify our method. In addition, we measure the flux of the 2021 Perseid outburst.

\section{Methods} \label{sec:methods}

\subsection{Global Meteor Network Hardware and Data Reduction}

The Global Meteor Network (GMN) is an amateur-professional collaboration project with over 600 video meteor cameras in more than 30 countries \citep{vida2021global}. The GMN uses low-cost consumer-grade security cameras which are a good match to the sensitivity and field of view of a human observer. In the most common configuration, a Sony IMX291 CMOS sensor is paired with a 3.6~mm f/0.95 lens to provide a field of view of $88^{\circ}\times48^{\circ}$ and a stellar limiting magnitude of $+6.0 \pm 0.5^{\mathrm{M}}$ at 25 frames per second, though other lens configurations are occasionally used. 

The camera is controlled via a Raspberry Pi 4 single-board computer (used currently, circa 2022). This runs the fully automated Raspberry pi Meteor Station (RMS) software\footnote{RMS software: \url{https://github.com/CroatianMeteorNetwork/RMS}} which schedules observations, captures the video stream, detects meteors, and calibrates the data \citep{vida2016open}. The final data product at each station is a list of time, position (right ascension, declination) and magnitude measurements for every point on the meteor's trajectory. These observations are uploaded to a central server which correlates them with other stations and computes trajectories and orbits \citep{vida2020estimating}.

To conserve disk space, RMS uses a rolling Four-frame Temporal Pixel video compression scheme \citep{gural2011california}. This method assumes that meteors are the brightest objects visible in a sub-area of the frame for a period on the order of seconds. The algorithm then compresses blocks of 256 video frames into 4 images: a per-pixel maximum stack, an image where the frame index of pixel maximum occurrence is encoded as pixel level, a per-pixel average stack, and per-pixel standard deviation image \citep{vida2021global}. At the normal GMN frame rate of 25 frames per second each 256-frame block has a duration of 10.24 seconds.

As a first step, a star extraction algorithm identifies stars on the average pixel stack and fits a Gaussian point spread function on each candidate star \citep{vida2016open}. The full width at half maximum (FWHM) is determined for every detected star. If more than 20 stars are detected (which from examination of many camera images had been determined to indicate that at least a part of the sky is clear), the meteor detection algorithm is run, producing measurements of meteor position and brightness.

When a camera is first set up, an initial astrometric and photometric calibration is manually produced to ensure quality. For the most common 3.6~mm lens configurations, the average astrometric accuracy is $\sim1$~arcmin, and the photometric uncertainty is $\pm 0.2^{\mathrm{M}}$. However, the camera pointing drifts slightly throughout the night due to thermal expansion of the camera housing bracket, amounting to as much as 10 arcmin. The sky conditions such as transparency also change, so an automated recalibration procedure is run for every detected meteor. In the recalibration step, the lens distortion and vignetting parameters are kept fixed, and only the pointing and the photometric offset are fit \citep{vida2021global}.

\subsection{Shower Association}

Single-station shower association is performed by first fitting a great circle to the observed equatorial coordinates of the meteor \citep[see Section 3.2 in][]{vida2020estimating}. If this great circle at any point is within 3$^{\circ}$ from the predicted position of the apparent radiant of a meteor shower, the meteor is tentatively associated with the shower. 

In this first association step, the geocentric radiants of all possible meteor showers occurring at the time of the meteor event are computed by applying empirically derived radiant drifts based on the shower list from \citet{vida2021global}. The equatorial coordinates of the apparent radiant ($\alpha_a, \delta_a$) are numerically inverted assuming a mean meteor height of 100~km and the geographical coordinates of the observer (i.e. the camera). From the great circle fit of the meteor observation, the normal vector $\mathrm{\textbf{n}}$ describing the orientation of the great circle (in equatorial coordinates) can be simply computed as:

\begin{align}
    \hat{\mathrm{\textbf{n}}} = \frac{\mathrm{\textbf{n}}}{|\mathrm{\textbf{n}}|} \,, \\
    \alpha_n = \arctantwo ( \hat{n}_y, \hat{n}_x) \,, \\
    \delta_n = \arcsin \hat{n}_z \,.
\end{align}

The closest radiant approach distance, which should be $< 3^{\circ}$ for a tentative association, is the angular separation between the normal and the apparent shower radiant offset by $\pi/2$:

\begin{align}
    \theta_{\mathrm{rad}} = |\pi/2 - \arccos(\sin \delta_n \sin \delta_a + \cos \delta_n \cos \delta_a \cos(\alpha_n - \alpha_a))|  \,.
\end{align}

A second check is made ensuring that the meteor is coming from the radiant instead of going towards it. This is simply achieved by confirming that the angular separations between the apparent radiant and the meteor begin point is smaller than the angular separation between the radiant and the end point.

\begin{figure*}
	\begin{center}
		\includegraphics[width=\linewidth]{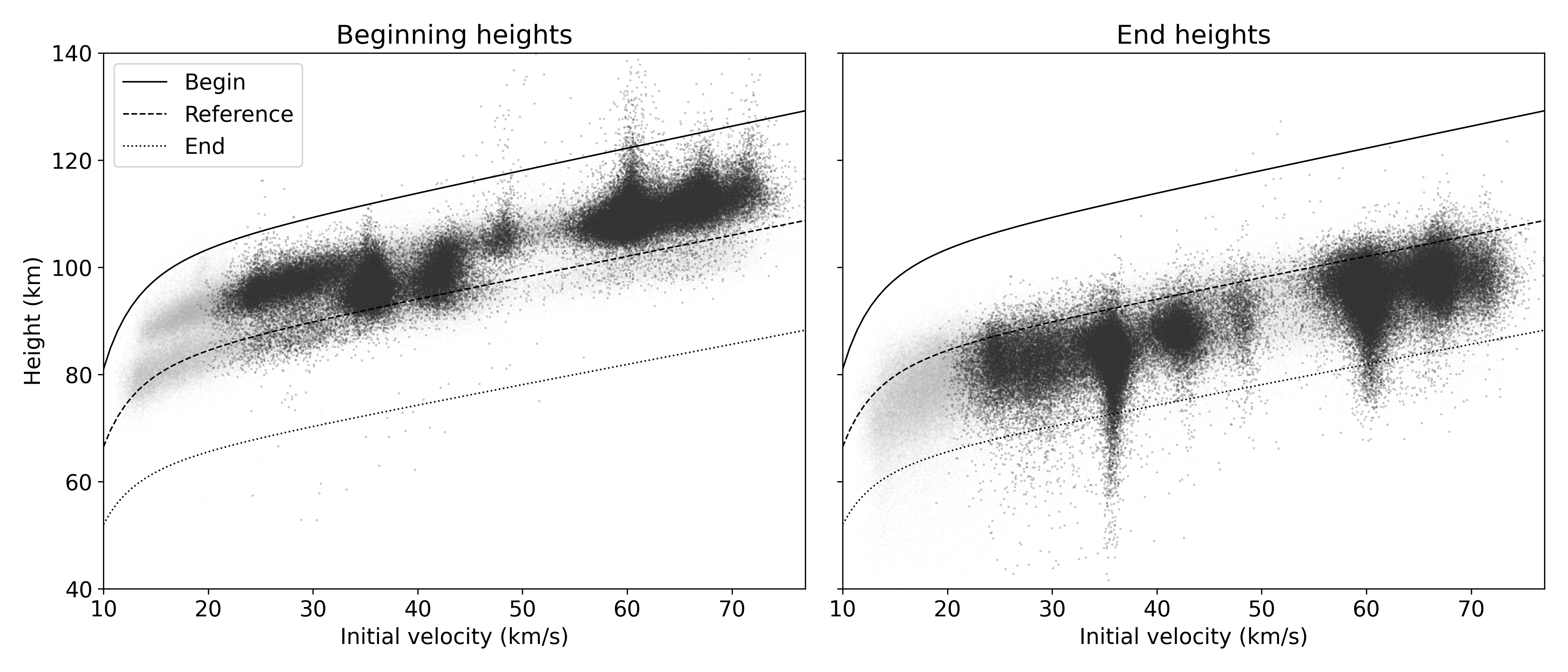}
	\end{center}
	\caption{Comparison between begin (left) and end heights (right) observed for complete lightcurves from multi-station GMN data. Light grey dots are all meteors, and dark grey are only meteors from major showers of interest as specified in \protect\citet{moorhead2019meteor}. The adopted begin height for single-station shower association is shown with a solid line (top), the reference height at which collection areas are computed is shown with a dashed line (middle), and the end for shower association with a dotted line (bottom). The populations where excursions are seen above the line (near 60~km/s for begin heights) or below for end heights (at 35 and 60~km/s) correspond to shower fireballs which are insignificant in flux calculations.}
	\label{fig:ht_model}
\end{figure*}

Finally, the meteor is definitively associated with a shower if its observed angular velocity is consistent with the known shower radiant/speed and the corresponding begin and end heights are in the expected range of heights \citep{Duffy1987}. Meteor heights are known to vary with speed \citep{jenniskens2016cams} and meteoroid type \citep{vida2018modelling}; at typical meteoroid sizes the GMN observes, asteroidal meteoroids penetrate about 5~km deeper than cometary meteoroids. To derive heights for various showers, we fit an empirical model to all observed begin and end heights (in kilometers) using the GMN trajectory database \citep{vida2021global}:


\begin{align}
    H_{\mathrm{beg}} = 97.84 + 0.41 v_0 - 20919.39 v_0^{-3} \,, \label{eq:meteor_ht_beg} \\
    H_{\mathrm{end}} = 59.48 + 0.37 v_0 - 11193.74 v_0^{-3} \label{eq:meteor_ht_end} \,,
\end{align}

\noindent where $v_0$ is the initial in-atmosphere velocity in km/s. The shape of the model closely follows the bulk of the observations, but has been shifted such that 99\% of observed meteors are within the empirical height range. The empirical fits compared to data are shown in Fig. \ref{fig:ht_model}.

Three pieces of information are available to compute the observed range of heights assuming a shower membership: the unit vector pointing to the meteor beginning in horizontal coordinates $\hat{\mathrm{\textbf{v}}}_b$, the vector pointing to the end $\hat{\mathrm{\textbf{v}}}_e$, and the total observed length of the meteor $d$. The total observed length in this construct is simply the product of the observed duration and the known shower velocity. Vectors $\hat{\mathrm{\textbf{v}}}$ are computed by taking the observed azimuths $\Lambda$ and altitudes $\epsilon$ of the beginning and the end of the meteor and separating them into Cartesian components:

\begin{align}
    \mathrm{v}_x = \cos \epsilon \cos \Lambda \,, \\
    \mathrm{v}_y = \cos \epsilon \sin \Lambda \,, \\
    \mathrm{v}_z = \sin \epsilon \,.
\end{align}

\noindent The same is done using the apparent horizontal coordinates of the radiant, resulting in a meteor shower radiant vector $\hat{\mathrm{\textbf{v}}}_r$.

\begin{figure}
	\begin{center}
		\includegraphics[width=\linewidth]{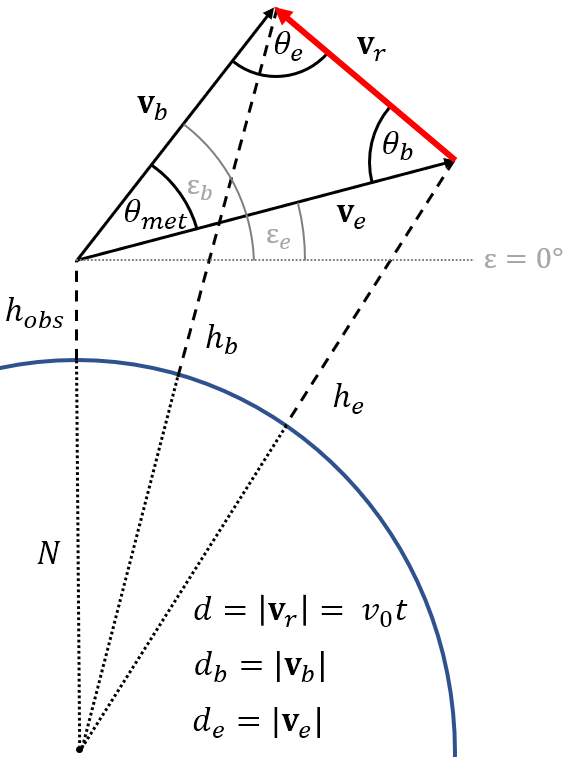}
	\end{center}
	\caption{Diagram showing how the observed begin and end heights are estimated. The red arrow is the meteor pointing in reverse, towards the radiant.}
	\label{fig:height_diagram}
\end{figure}

Taking the curvature of the Earth into account, the observed meteor begin height $h_{\mathrm{b}}$ and the end height $h_{\mathrm{e}}$ are computed as follows:

\begin{align}
    \theta_{\mathrm{met}} = \arccos \left ( \hat{\mathrm{\textbf{v}}}_b \cdot \hat{\mathrm{\textbf{v}}}_e \right ) \,, \\
    \theta_{\mathrm{b}} = \arccos \left ( \hat{\mathrm{\textbf{v}}}_r \cdot (-\hat{\mathrm{\textbf{v}}}_e) \right) \,, \\
    \theta_{\mathrm{e}} = \arccos \left ( -\hat{\mathrm{\textbf{v}}}_r \cdot (-\hat{\mathrm{\textbf{v}}}_b) \right) \,, \\
    d_b = d \frac{\sin \theta_{\mathrm{b}}}{ \sin \theta_{\mathrm{met}}} \,, \\
    d_e = d \frac{\sin \theta_{\mathrm{e}}}{ \sin \theta_{\mathrm{met}}} \,, \\
    R_{\mathrm{obs}} = N + h_{\mathrm{obs}} \,, \\
    h_{\mathrm{b}} = \left | -N + \sqrt{ d_b^2 + R_{\mathrm{obs}}^2 - 2 d_b R_{\mathrm{obs}} \cos (\epsilon_b + \pi/2 )} \right | \,, \\
    h_{\mathrm{e}} = \left | -N + \sqrt{ d_e^2 + R_{\mathrm{obs}}^2 - 2 d_e R_{\mathrm{obs}} \cos (\epsilon_e + \pi/2 )} \right | \,,
\end{align}

\noindent where $\theta_{\mathrm{met}}$ is the observed angular extent of the meteor in the sky, $\theta_{\mathrm{b}}$ the observer-end-begin angle, $\theta_{\mathrm{e}}$ the observer-begin-end angle, $d_b$ and $d_e$ are the linear distances from the observer to the begin and end point, respectively, $N$ is the distance between the centre of the Earth and surface at latitude $\varphi$ \citep[see Appendix D in][]{vida2020estimating}, $h_{\mathrm{obs}}$ is the elevation of the observer above the WGS84 ellipsoid, $R_{\mathrm{obs}}$ is the distance between the centre of the Earth and the observer, and $\epsilon_b$ and $\epsilon_e$ are the observed altitude of the meteor begin and end points above the horizon, respectively. Fig. \ref{fig:height_diagram} shows an illustration explaining the geometry.

If $H_{\mathrm{beg}} > (h_{\mathrm{b}} + h_{\mathrm{e}})/2 > H_{\mathrm{end}}$, the meteor is considered to be associated with the shower. This method is very sensitive to the estimated length of the meteor $d$, and may cause short meteors not to be associated at small distances from the radiant. We perform additional checks with the duration varied by $\pm 1$ video frame - if any of these instances satisfy the height criterion, the association is accepted. Potential issues this method might encounter with meteors that have low entry angles are avoided by imposing a minimum apparent radiant elevation limit, as described below in Section \ref{sec:corrections}.

\begin{figure*}
	\begin{center}
		\includegraphics[width=\linewidth]{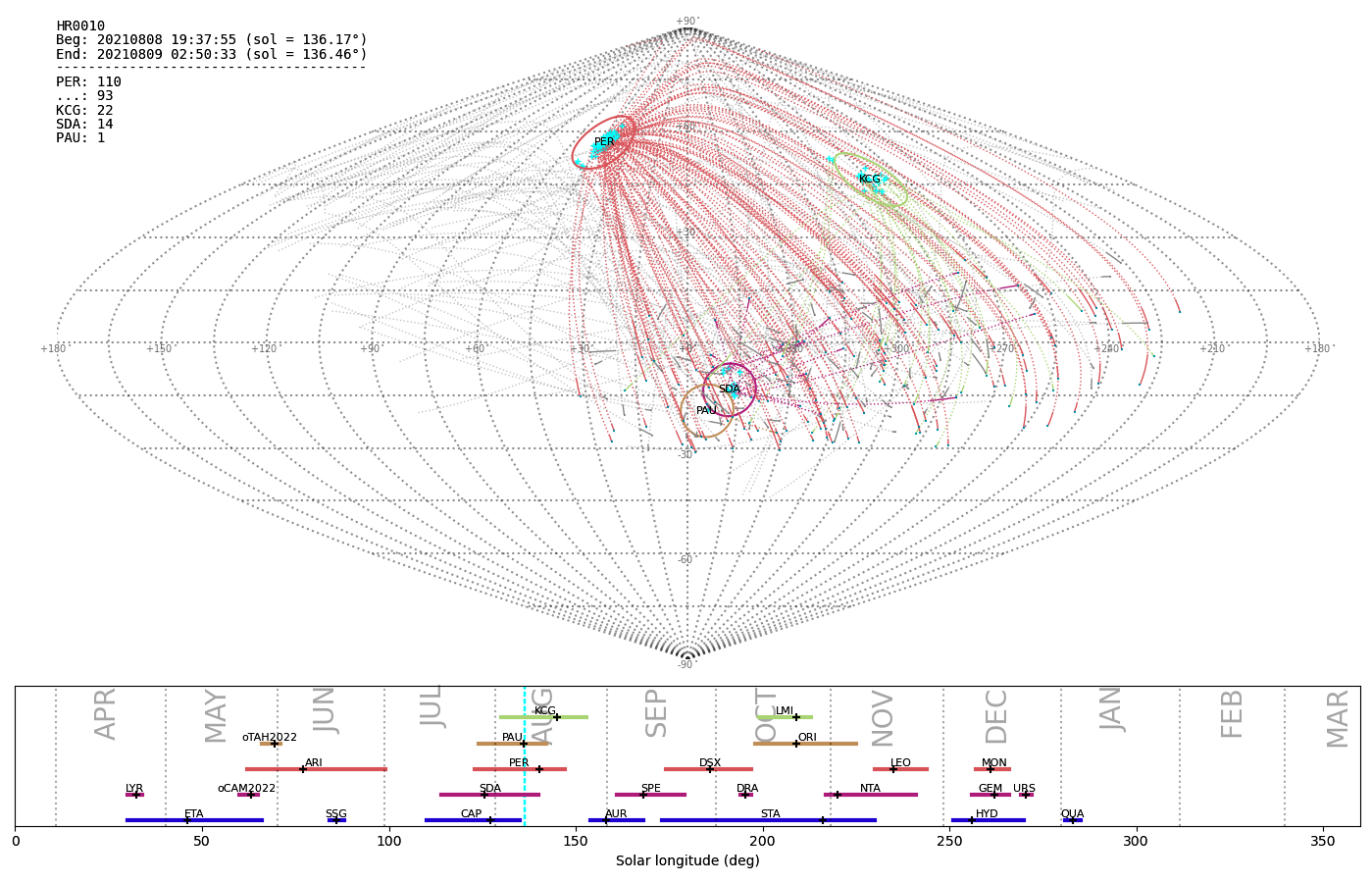}
	\end{center}
	\caption{Top: Individual meteor paths (solid lines ending with cyan circle) and their partial great circle projections (dotted lines, colored per shower named following the IAU three letter designation) for single-station shower association in the night of August 8-9, 2021 from the HR0010 station. All-sky map (sinusoidal project) is in equatorial coordinates showing mean locations of apparent shower radiants. The top legend shows the time period of the observation and the number of single-station shower associations as well as sporadics (given by ``..."). The point of closest approach for each great circle to the radiant is designated by a cyan cross. Bottom: A list of the annual showers based on the International Astronomical Union shower catalogue considered for flux computation and their activity periods \protect\citep{moorhead2019meteor}.}
	\label{fig:PER_shower_association}
\end{figure*}

Fig. \ref{fig:PER_shower_association} is an example of an all-sky map in equatorial coordinates which shows the paths of observed meteors from one camera on one night, and the associated shower membership using our criteria.

\subsection{Determining the Effective Period of Camera Observation} \label{subsec:observing_period}

The first step in measuring the flux for a given camera is choosing observing periods during which the measurements will be made. We define a favourable observing period as a time when at least 50\% of catalog stars that are predicted to be observed (based on known pointing and expected sensitivity of the camera) are registered by the RMS star detector. The method takes into account lens vignetting, atmospheric extinction, the changing stellar limiting magnitude, the presence of the Moon, and masked out portions of the field of view due to the presence of obstructions.

\begin{figure}
	\begin{center}
		\includegraphics[width=\linewidth]{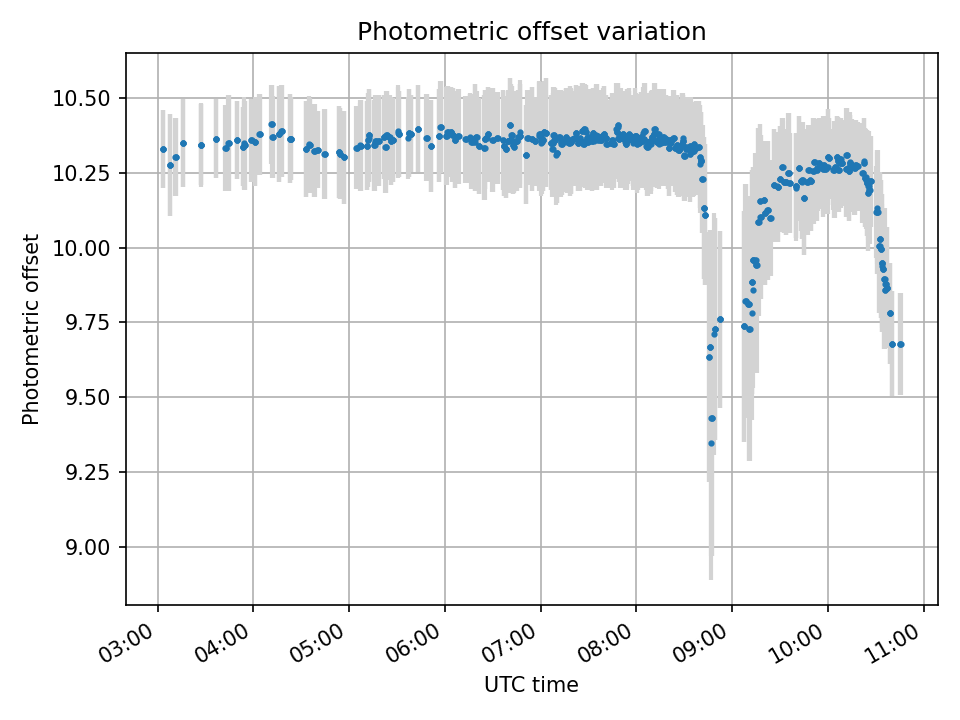}
	\end{center}
	\caption{Variation of the photometric offset (in units of magnitude) as observed on August 14, 2021 by a camera in South Dakota, USA. The dip around 9:00 UTC corresponds to a cloudy period, and the error bars indicate $1\sigma$ fit errors.}
	\label{fig:p0_variation}
\end{figure}

All observations when the Moon was at least 25\% illuminated and inside of the field of view are rejected. The astrometry and photometry recalibration is performed in 5 minute steps throughout the night to determine the limiting stellar magnitude. This recalibration procedure produces measurements of the photometric offset $p_0$ which is used to convert instrumental observations of pixel intensity to magnitudes using the classical relation \citep{Hawkes2002}:

\begin{align}
    M = -2.5 \log_{10} ( S_{\mathrm{px}} ) + p_0 \,,
\end{align}

\noindent where $M$ is the magnitude and $S_{\mathrm{px}}$ is the sum of pixel intensities. The effect of vignetting and atmospheric extinction is also removed, as described in \citet{vida2021global}. The variation in the photometric offset due to changing sky conditions can be significant, up to one magnitude during the course of the night, as shown in Fig. \ref{fig:p0_variation}.

The stellar limiting magnitude $M_s$ is computed using the photometric offset and the following empirical relation:

\begin{align} \label{eq:stellar_lm}
    M_s = 0.83 p_0 - 2.59 \,.     
\end{align}

\begin{figure*}
	\begin{center}
		\includegraphics[width=\textwidth]{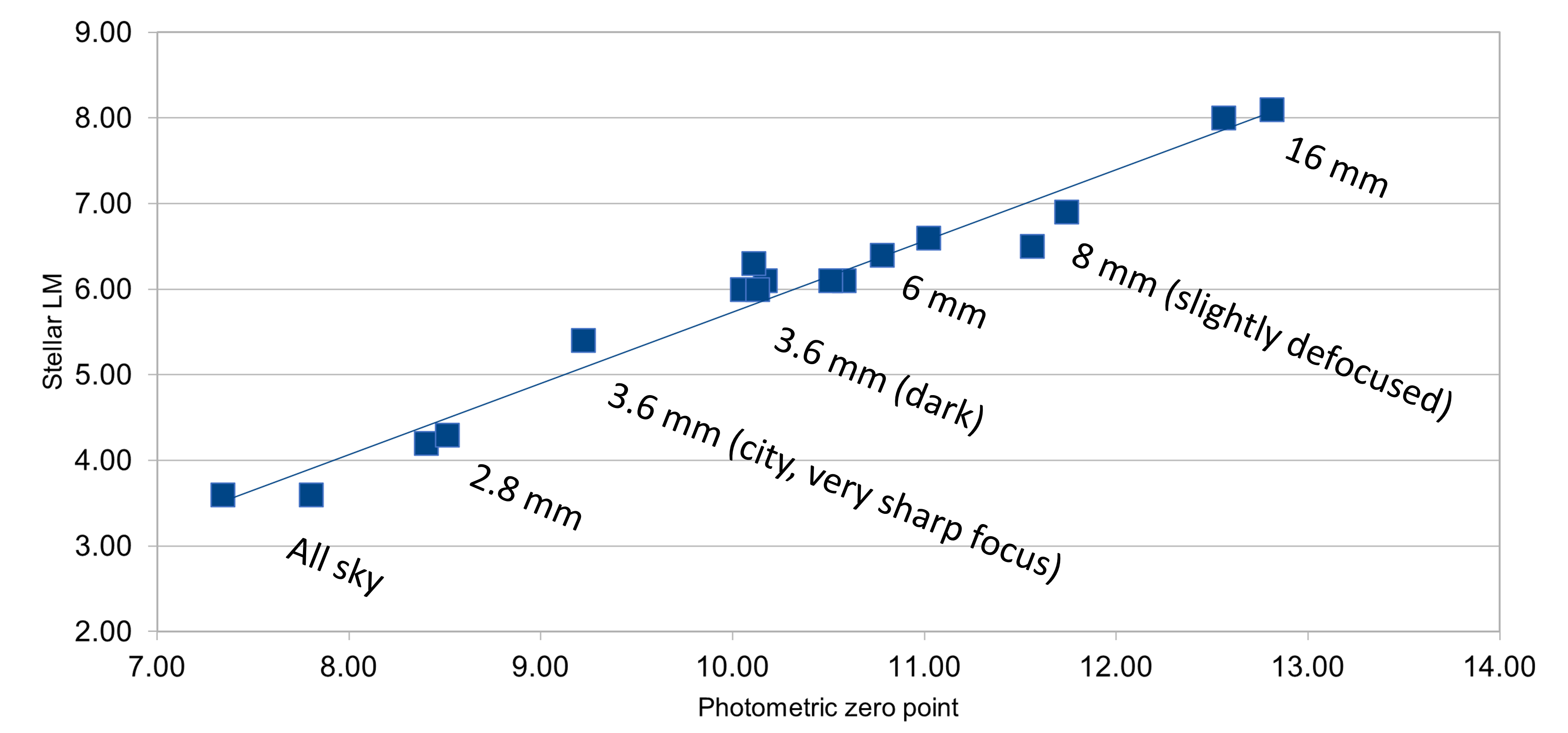}
	\end{center}
	\caption{Manually determined empirical fit of stellar limiting magnitude measurements as a function of the photometric offset. The focal lengths of the different GMN systems used for the fit are indicated for every group of points.}
	\label{fig:lim_mag}
\end{figure*}

This relation was derived via manually estimated limiting stellar magnitude measurements covering the whole range of GMN systems, from all sky to narrow field lenses, as shown in Fig. \ref{fig:lim_mag}. The manual procedure included increasing the stellar limiting magnitude of catalog stars in increments of $0.1^{\mathrm{M}}$ until the stars in the image could not be visually determined to match the catalog stars. This procedure closely mimics limiting magnitude estimation during visual meteor observations \citep{rendtel2020handbook}. Moreover, it accounts for all end-to-end biases (except those from the RMS software) such as photometric processing by the camera, noise etc. The linear empirical fit has a standard deviation of $\pm 0.25^{\mathrm{M}}$. 


Unlike other source-extraction software, the RMS star detector was designed to robustly and quickly detect around 50-100 stars during clear skies, even though more stars can be seen in the image. By comparing the predicted limiting magnitude from our empirical relation to the limiting magnitude recovered by the software, we determined that the detection threshold of the software translates into a $1.2^{\mathrm{M}}$ brighter limiting magnitude than is given in Eq. \ref{eq:stellar_lm}.

The GAIA DR2 catalog \citep{gaia2018dr2} is queried for stars up to the instrumental limiting magnitude and the detected stars are matched to the catalog. This gives the number of matched stars $N_{\mathrm{matched}}$. The edge of the image is sampled with 4 vertices on each side to capture the effects of lens distortion. The list of catalog stars is filtered such that only the stars inside the polygon in equatorial coordinates described by the 16 vertices are taken. Stars whose apparent magnitude (with vignetting and atmospheric extinction applied) is fainter than the limiting magnitude are removed. The celestial coordinates are mapped into image coordinates and stars in masked regions are rejected. The number of remaining stars is thus the number of predicted stars $N_{\mathrm{predicted}}$ which a camera should detect under good conditions. 


To determine when conditions are suitable for inclusion in flux estimation, the ratio $N_{\mathrm{matched}}/N_{\mathrm{predicted}}$ is considered. By investigating observations from cameras in various conditions, we found empirically that a ratio of 0.5 accurately discriminates between acceptable and overcast skies. This method works regardless of observing conditions, sky quality, and lens used, as long as there is a good initial calibration. It also identifies poorly calibrated cameras, as the algorithm will not match any stars and will thus always produce a ratio of 0.

\begin{figure}
	\begin{center}
		\includegraphics[width=\linewidth]{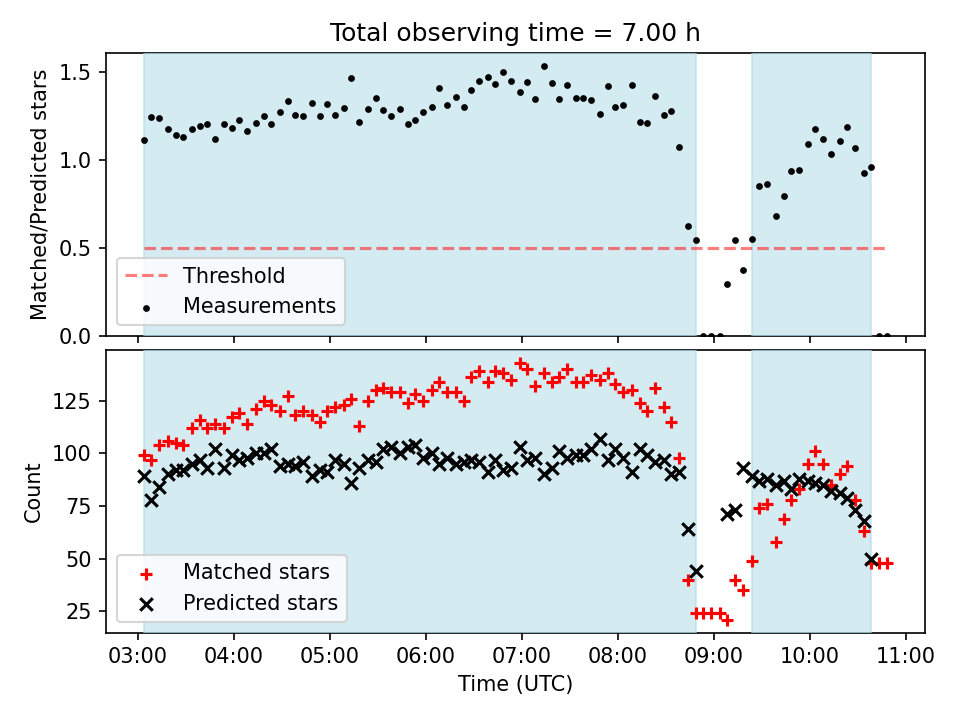}
	\end{center}
	\caption{Observing periods determined for a camera in South Dakota on August 14, 2021. Blue shading indicates the observing periods when the ratio was above the threshold for at least 90 minutes. The gap around 9:00 UTC corresponds to a cloudy period which would not be used for flux measurements.}
	\label{fig:observing_periods}
\end{figure}

Observing periods are included for flux estimation only for cameras where the ratio is above the threshold for at least 90 minutes with no more than 15 minutes per 90 minute block below the threshold. Fig. \ref{fig:observing_periods} shows an example of the estimation of good observing periods for one night using this approach. 

We tested the method on dozens of nights around the peaks of investigated showers. The automated results produced cloudy time windows that were within five minutes in start/stop time compared to manual estimates. Were the algorithm to produce erroneous results on some nights that were not covered by the verification procedure, it could only do so conservatively. The requirement that at least 20 catalog stars are matched to observed stars (all within 0.5~px for at least 90 minutes) makes it very unlikely for the algorithm to classify a cloudy night as acceptable.

\subsection{Computing Meteoroid Flux}

The flux of meteors in the Earth's atmosphere is defined as the number of meteors passing through a unit of area in a unit of time, down to a certain mass limit:

\begin{align} \label{eq:flux}
    F(> m_{\mathrm{lim}}) = \frac{N_{\mathrm{meteors}}}{A_e t} \,,
\end{align}

\noindent where $N_{\mathrm{meteors}}$ is the number of meteors of mass larger than $m_{\mathrm{lim}}$ observed over an area $A_e$ in time $t$. The $A_e t$ term in the denominator is often referred to as the time-area product (TAP) \citep{musci2012optical}. Conventionally, the flux is expressed in the units of meteoroids~km$^{-2}$~h$^{-1}$ \citep{brown1998}. For convenience, the TAP can be expressed in units of meteoroids per 1000~km$^2$~h, as most major showers have their peak flux in the range of tens $\times 10^{-3}$ meteoroids~km$^{-2}$~h$^{-1}$.

For the GMN, the number of meteors is provided as-is by the meteor detector, the limiting mass is computed using an empirical conversion relating the mass-peak magnitude and velocity \citep{verniani1973analysis}, and the time is simply the duration of a chosen time bin. The flux is most sensitive to estimation of the effective collection area $A_e$  \citep[also called the ``reduced area'' in some literature;][]{koschack1990determination1} which includes all observational biases.

\subsubsection{Raw Atmospheric Collecting Area}

The effective collection area is computed by starting with the true area of the atmosphere observed by the camera at meteor heights (assumed to be a segment of a spherical shell at a height mid-way between the start and end height as a function of speed from Fig \ref{fig:ht_model}, at around 100~km), and correcting for observational biases. The effective area is always smaller than the true area, as observational biases decrease the number of observed meteors compared to ideal conditions. The bias corrections include terms for the camera sensitivity, meteor distance, radiant elevation, and meteor angular velocity. All of these values change across the field of view and over time.

The observing period determined in Section \ref{subsec:observing_period} is binned into 5 minute intervals, which was chosen as the smallest unit of time likely to be needed for flux calculations in most circumstances. For every time bin, the field of view is divided into $20 \times 11$ rectangular blocks (for 16:9 sensor formats) and the correction variation is applied per block across the field of view. The outline of the blocks are projected to meteor heights from which raw atmospheric collecting areas are computed.

To compute the area, the geographical coordinates of the four corners of the block at the reference meteor height are needed. As a first step, apparent azimuth $\Lambda$ and altitude $\epsilon$ of the corners are computed using the astrometric plate, accounting for atmospheric refraction and lens distortion \citep{vida2021global}. Blocks with all four corners below an altitude of 20$^{\circ}$ are ignored. This avoids applying large corrections due to high meteor range and atmospheric extinction for large atmospheric areas. We take this conservative approach to avoid a major source of uncertainty \citep{vaubaillon2021malbec}, namely high corrections which may skew absolute estimates when debiasing \citep{galligan2004orbital}. As part of this procedure we also remove all meteors observed below an altitude of $\epsilon < 20^{\circ}$.

Given the known WGS84 geographical coordinates of the observer (latitude $\varphi$, longitude $\lambda$, elevation $h_{\mathrm{obs}}$), the range $r$ between the observer and a corner line of sight projected at the meteor height $h_{\mathrm{met}}$ can be computed with the curvature of the Earth taken into account as:

\begin{align}
    d_s = N + h_{\mathrm{obs}} \,, \\
    d_m = N + h_{\mathrm{met}} \,, \\
    \beta = \epsilon + \arcsin \frac{d_s \cos \epsilon}{d_m} \,, \\
    r = d_m \frac{\cos \beta}{\cos \epsilon} \,, \label{eq:range}
\end{align}

\noindent where $N$ is the distance between the centre of the Earth and surface at latitude $\varphi$ \citep[see Appendix D in][]{vida2020estimating}, $\beta$ is the angle between the observer, the centre of the Earth, and a point at height $h_{\mathrm{met}}$ which lies on the corner line. 

With the range $r$ known, the Earth-centered Earth-fixed (ECEF) coordinates of a corner point can be computed as:

\begin{align}
    S = -r \cos \epsilon \cos \Lambda \,, \\
    E =  r \cos \epsilon \sin \Lambda \,, \\
    Z =  r \sin \epsilon \,, \\
    x = x_{\mathrm{obs}} + S \sin \varphi \cos \lambda - E \sin \lambda + Z \cos \varphi \cos \lambda \,, \\
    y = y_{\mathrm{obs}} + S \sin \varphi \sin \lambda +  E \cos \lambda + Z \cos \varphi \sin \lambda \,, \\
    z = z_{\mathrm{obs}} - S \cos \varphi + Z \sin \varphi \,, 
\end{align}

\noindent where $S, E, Z$ are South, East, and vertical components of the range vector, $x_{\mathrm{obs}}, y_{\mathrm{obs}}, z_{\mathrm{obs}}$ are ECEF coordinates of the observer \citep[see Appendix D1 in][]{vida2020estimating}, and $x, y, z$ are ECEF coordinates of the corner point at the meteor height. The final step is to convert the resulting ECEF coordinates into geographical coordinates \citep[see Appendix D2 in][take apparent sidereal time at Greenwich $\theta'_0 = 0$ to get ECEF instead of Earth-centered inertial coordinates]{vida2020estimating}.

We compute the total area $A$ bound by the four sets of geographical coordinates ($\varphi_i$, $\lambda_i$) by performing a line integral on a sphere (the radius of the sphere is assumed to be $N + h_{\mathrm{met}}$). The Green's theorem allows the integration to be split into piece-wise segments bound by each side of the polygon\footnote{The equations were first derived for the \texttt{sphericalgeometry} library; we found no other published literature on the subject: \url{https://github.com/anutkk/sphericalgeometry}, accessed on 2022/04/05.}:

\begin{align}
    \chi_i = \sin^2 \left ( \frac{\varphi_i}{2} \right) + \cos \varphi_i \sin^2 \left ( \frac{\lambda_i}{2} \right) \,, 
\end{align}
\begin{align}
    \vartheta_i = 2 \arctantwo \left ( \sqrt{\chi_i}, \sqrt{1 - \chi_i} \right ) \,, 
\end{align}
\begin{align}
    \bar{\vartheta}_i = \vartheta_i + (\vartheta_i - \vartheta_{i+1})/2  \qquad \forall i \in [1, 4] \,, 
\end{align}
\begin{align}
    \gamma'_i = \arctantwo \left ( \cos \varphi_i \sin \lambda_i, \sin \varphi_i \right ) \bmod 2 \pi \,, 
\end{align}
\begin{align}
    \Delta \gamma'_i = \left ( (\gamma'_i - \gamma'_{i+1} + \pi ) \bmod 2 \pi \right) - \pi  \qquad  \forall i \in [1, 4] \,, 
\end{align}
\begin{align}
    A' = \sum_{i=1}^4 \Delta \gamma'_i \left ( 1 - \cos \bar{\vartheta}_i \right ) \,, 
\end{align}
\begin{align}
    A = \min \left (A', 1 - A' \right) (N + h_{\mathrm{met}})^2 \,.
\end{align}

\noindent Note that the input vector of latitudes and longitudes has five elements - the first and the last element are the same to form a closed polygon. To satisfy the Green's theorem, the order of the vertices needs to be counter-clockwise.

\begin{figure}
	\begin{center}
		\includegraphics[width=\linewidth]{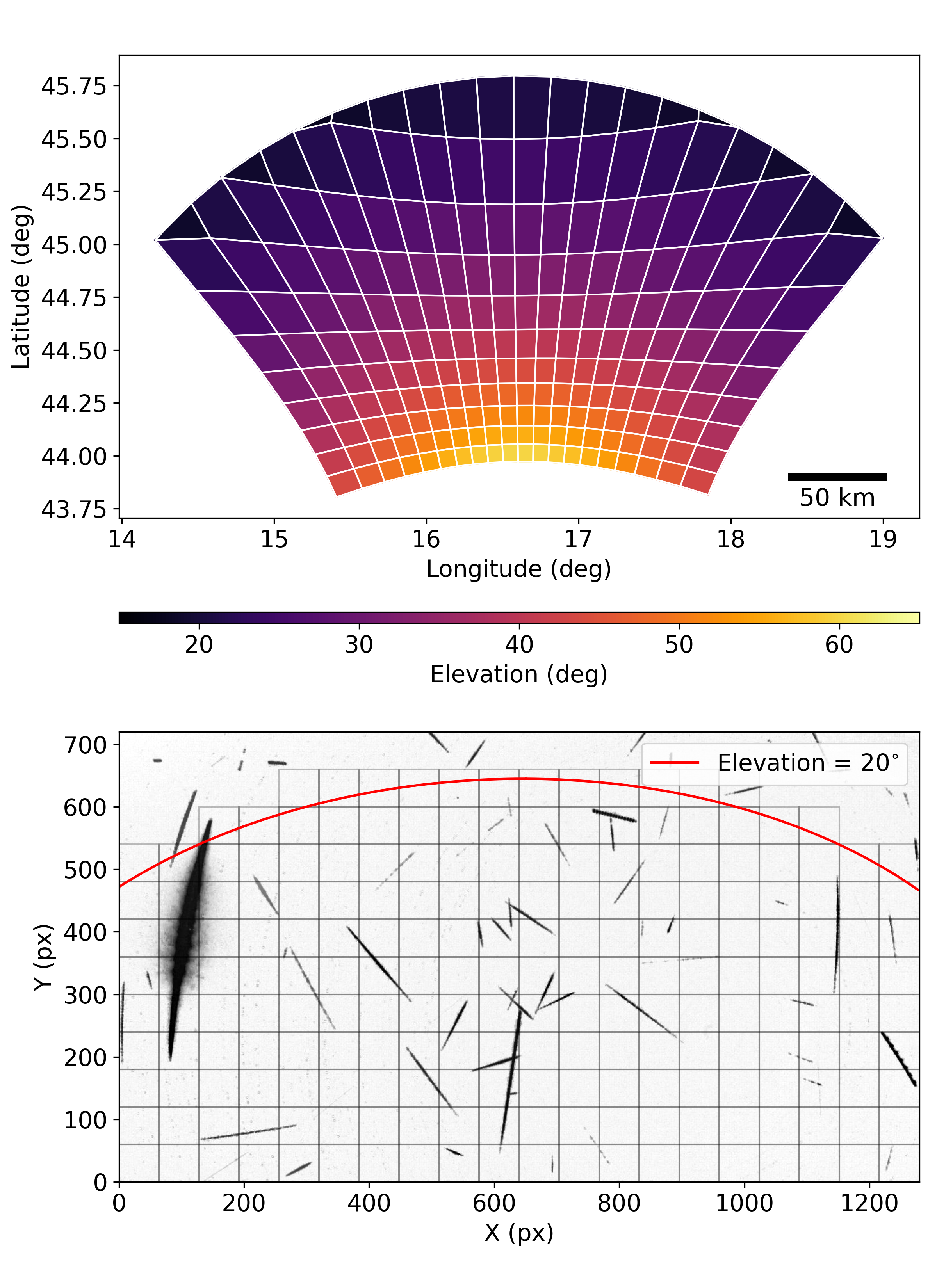}
	\end{center}
	\caption{Top: Blocks plane projected to the Earth's surface from the station HR000K at the height of 100~km, with the apparent elevation color coded. Blocks pointing to lower apparent elevations are significantly larger in area than those near the zenith, but they also have larger bias corrections. Bottom: Co-added images including meteors as detected from the station HR000K. The blocks used for area estimation are overlaid on the image, and the $20^{\circ}$ elevation limit is plotted as a red line. The colors are inverted, and the image is upside down so that the orientation matches the top inset.}
	\label{fig:fov_grid}
\end{figure}

With the per-block raw area computed, the fraction of the image that is masked out in the given block (due to the presence of local obstructions near the camera) is removed from the computed area. This final step yields the total raw area per block observed by the camera (excluding any observational biases). This area remains constant over time, as long as the camera pointing does not change. Fig. \ref{fig:fov_grid} shows an example of an image divided into blocks and the associated projection into geographical coordinates at the height of 100~km.

\subsubsection{Correcting for Observational Biases} \label{sec:corrections}

Prior to introducing any corrections, area blocks which are within 15$^{\circ}$ from the radiant are excluded, as are observation periods when the radiant was within 15$^{\circ}$ of the middle of the field of view. This is done to avoid the radiant singularity and unphysically large corrections as the apparent meteor angular velocity is zero at the radiant. Additionally, all observation periods when the apparent radiant elevation was below 15$^{\circ}$ are rejected. These limits were chosen such that no corrections exceed a factor of 10, minimizing potential systematic errors due to modelling assumptions, as adopted in earlier meteor debiasing work \citep{galligan2004orbital}.


For every time bin, a reference meteor limiting magnitude $M_m$ in the middle of the focal plane is computed following \citet{blaauw2016optical}:

\begin{align}
    M_m = M_s - 5 \log_{10} \left( \frac{r_{\mathrm{mid}}}{100} \right) - 2.5 \log_{10}  \left ( \frac{v_0 F \sin \theta }{\mathrm{{FPS} ~r_{\mathrm{mid}}~\mathrm{FWHM} }} \right) \,,
\end{align}

\noindent where $r_{mid}$ is the range in kilometers from the centre of the focal plane to a point at the meteor height (computed using Eq. \ref{eq:range}), $v_0$ is the in-atmosphere meteor velocity in km/s, $F$ is the plate scale in px/deg, FPS is the frame rate, and FWHM is the full width at half maximum of the point spread function in px. The FWHM is determined by fitting a Gaussian point spread function on individual detected stars every 5 minutes and taking a median value. The meteor limiting magnitude $M_m$ varies across the field of view; however the per-block corrections given below even out this variation so that only one reference $M_m$ need be used at any time.

The collection area of a meteor shower is estimated at the reference height at which meteors are observed (see Fig. \ref{fig:ht_model}). The reference height is simply the mean of the model begin and end height, given in Eq. \ref{eq:meteor_ht_beg} and \ref{eq:meteor_ht_end}. The reference height can also be manually specified. Because meteor showers contain meteoroids of different masses, the heights can vary by a few kilometers. From GMN meteor observations, we find that a standard deviation of $\sigma_h = 2$~km represents this variation well. To take this effect into account, we use a Gaussian weighting scheme. The distribution is centered on the estimated reference height and probabilities are computed with a height delta of 2~km between $\pm 3 \sigma_h$. The weights are normalized such that $\sum_{h=60}^{130} w(h) = 1$, where the limits corresponds to the model limits.

The total effective collection area $A_e$ at the given time is computed as:

\begin{align}
    A_e = \sum_{h=60}^{130} \sum_{x, y} w(h) A(x, y, h) \left ( c_{\mathrm{sensitivity}} c_{\mathrm{range}} c_{\epsilon} c_{\theta} \right)^{s - 1} \,,
\end{align}

\noindent where $w(h)$ is the weight at each height, $A(x, y, h)$ is the raw area of each block at the height $h$, the $c$ parameters are individual per-block corrections, and $s$ is the differential mass index of the shower. Note that this is physically the same basic approach as is used to compute meteor radar collecting areas \citep{Kaiser1960}.

The sensitivity correction takes into account optical vignetting and atmospheric extinction. The GMN photometric calibration models both of these effects \citep{vida2021global}. A no-atmosphere magnitude $M_{\mathrm{na}}$ is computed using the full model and an arbitrary pixel level sum value $S_{\mathrm{px}} = 400$; this magnitude includes the vignetting and extinction corrections. The magnitude is then converted back into a pixel value without these corrections (simply $S_{\mathrm{raw}} = 10^{(M_{\mathrm{na}} - p_0)/(-2.5)}$ recalling that p$_0$ is the photometric offset). The total sensitivity correction is thus:

\begin{align}
c_{\mathrm{sensitivity}} = \frac{ S_{\mathrm{px}} }{ S_{\mathrm{raw}} }  \,.
\end{align}

The range correction is:

\begin{align}
    c_{\mathrm{range}} = \left ( \frac{100}{r} \right )^2 \,,
\end{align}

\noindent where $r$ is the range in km along the middle of the block to the meteor height, as the meteor absolute magnitude is defined as the magnitude of a meteor seen from a range of 100 km.

The radiant elevation correction is:

\begin{align}
    c_{\epsilon} = \sin^{\gamma} \epsilon
\end{align}

\noindent where $\epsilon$ is the radiant elevation at the considered time, and $\gamma$ is the zenith exponent \citep{zvolankova1983dependence} which we keep fixed at 1.5 for all showers. $\gamma > 1$ indicates that there is a non-geometrical dependence on the observation efficiency with the radiant elevation, possibly due to atmosphere or ablation effects, and can range between 1 and 2. \citet{molau2013meteoroid} found that $\gamma$ varies from shower to shower; however they also found it very difficult to measure.

The angular velocity correction accounts for sensitivity loss due to the shortened effective exposure per pixel during a frame integration caused by the meteor angular velocity. Because of this, the meteor limiting magnitude varies across of the field of view. The correction normalizes the angular velocity to the limiting magnitude in the middle of the field of view. The per-block correction is:

\begin{align}
    c_{\theta} = \frac{ \theta_{\mathrm{block}} }{ \theta_{0} }  \,,
\end{align}

\noindent where $\theta_{\mathrm{block}}$ is the angular velocity in the middle of the considered block, and $\theta_{0}$ is the angular velocity in the middle of the focal plane. The angular velocity is computed as:

\begin{align}
    \theta = v_0 \frac{\sin \eta}{r} \,,
\end{align}

\noindent where $\eta$ is the angular distance from the radiant to the considered direction.

We ignore any correction for partial trails in the field of view as the GMN fields are large compared to typical apparent meteor path lengths that almost all meteors occur entirely inside the FOV. For narrower field of view this correction may need to be added \citep{blaauw2016optical}.

\subsection{Scaling the Flux to a Target Limiting Magnitude or Mass}

Optical observations of meteor showers are by definition brightness limited surveys. In this work, we use as reference sensitivity the meteor limiting magnitude in the middle of the field of view. However, as the limiting magnitude can change over time, an approach is required to scale observations to a standardized reference, which is traditionally taken as a meteor limiting magnitude of $+6.5^{\mathrm{M}}$ \citep{blaauw2016optical}. From an operational standpoint, however, it is helpful to have the sensitivity expressed as a limiting mass, as it will not depend on the spectral response of the system nor the velocity, so fluxes can be compared between different showers on the same mass scale.

From atmospheric observations \citep{vojavcek2019properties, vida2020new} and in-situ measurements of dust at parent comets \citep{fulle2016evolution}, it is known that masses of shower meteors are usually distributed according to a power-law distribution:

\begin{align}
    dN \propto m^{-s}dm \,,
\end{align}

\noindent where $dN$ is the number of meteoroids with mass in the interval $dm$ and $s$ is the differential mass index. The mass indices of showers can be measured using various methods \citep{koschack1990determination1,  blaauw2011meteoroid, pokorny2016reproducible, ehlert2020measuring, vida2020new}. Although it is possible to measure the mass index from the actual data set used to compute the flux \citep{molau2014obtaining}, in this work we use accepted values from the literature.

Traditionally, the meteor peak magnitude is taken as a proxy for mass \citep{jacchia1967}. From the magnitude-mass relation \citep{vida2020new}, it can be shown that the number of meteors of a given peak magnitude should follow:

\begin{align}
    r = \frac{N(M_m + 1)}{N(M_m)} \,,
\end{align}

\noindent where $r$ is the population index \citep{brown1996perseid} and $N(M_m)$ is the number of meteors in the integer magnitude bin $M_m$. Thus, to scale the flux from the observed meteor limiting magnitude $M_m$ to a reference limiting magnitude of $+6.5^{\mathrm{M}}$, the following relation can be used:

\begin{align} \label{eq:flux_scaling}
    F(6.5^\mathrm{M}) = F(M_m) r^{6.5 - M_m} \,.
\end{align}

\noindent To minimize the effect of the assumed mass index on the final result, a mean observed limiting magnitude can also be chosen as reference, in which case the factor of $6.5$ in the equation would be replaced with the limiting magnitude of choice.

Formally, from the definition of the magnitude, the conversion between the mass and the population index is \citep{ceplecha1998meteor}:

\begin{align} \label{eq:mass_index}
    s = 1 + 2.5 \log_{10} r \,.
\end{align}

\begin{figure}
	\begin{center}
		\includegraphics[width=\linewidth]{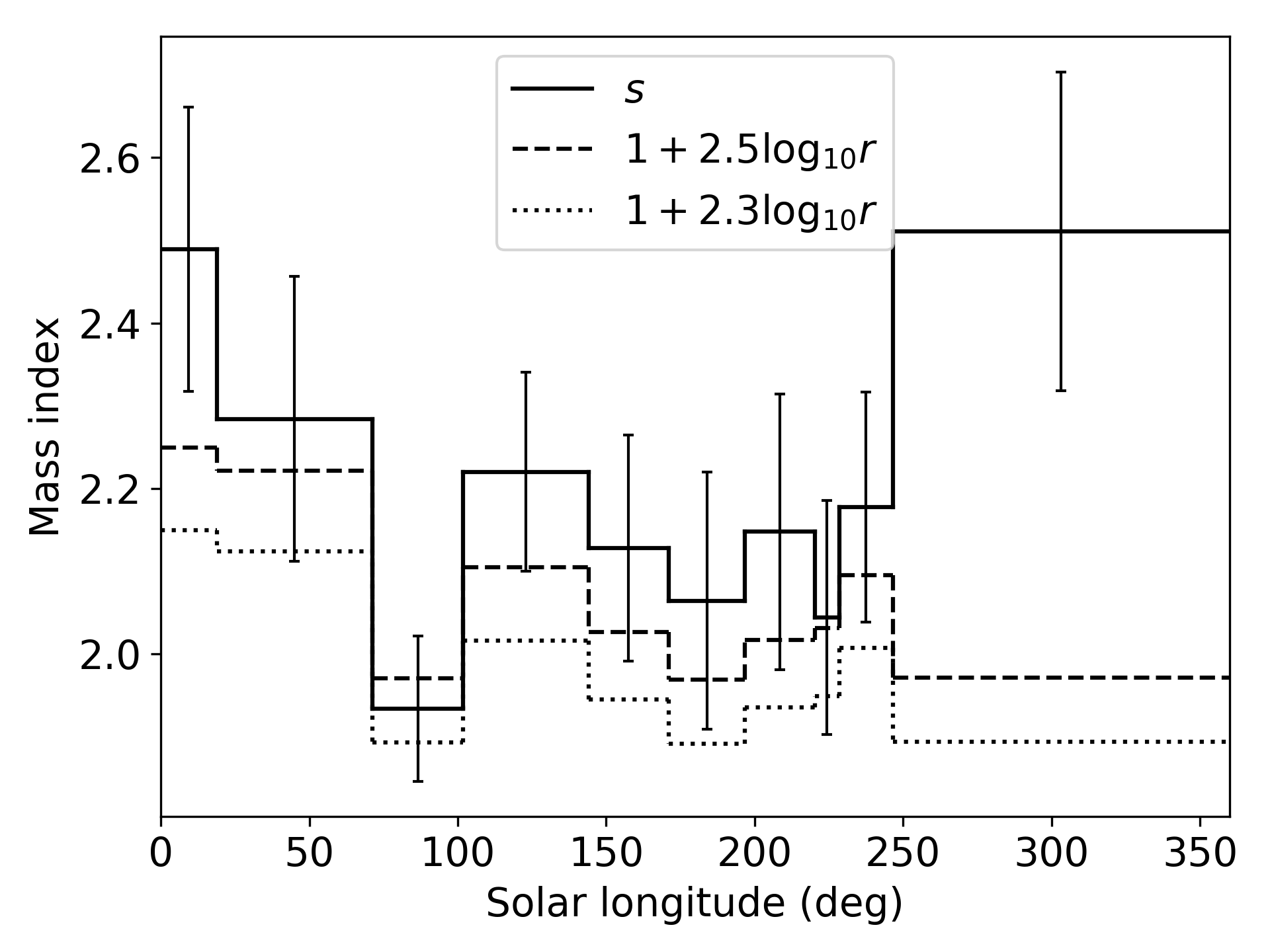}
	\end{center}
	\caption{Annual variation in the measured sporadic mass index (solid line, $1 \sigma$ errors shown) as derived using the Canadian Automated Meteor Observatory's influx camera using complete light curves to estimate mass \citep{vida2018first}. Every bin has a total of 250 meteors. The dashed line is the mass index derived from the population index using the factor of 2.5, and the dotted line is using the factor 2.3, all based on meteor peak magnitude only.}
	\label{fig:r_vs_s}
\end{figure}

\citet{verniani1973analysis} found that a factor of 2.3 better describes the conversion based on empirical observations of the meteor radar magnitude and mass. However, \citet{vida2020new} compared the directly measured population and mass indices using a robust statistical method on a high-quality manually reduced data set of optical meteors. They found that the conversion with the factor of 2.3 ($s = 1 + 2.3 \log_{10} r$) is not a good fit to the true mass index of sporadics. Of importance is that the mass index computed in that work was found from full integration of the complete light curves and conversion to total mass using a fixed luminous efficiency value, not just peak magnitudes. Here we repeat the analysis but also include the conversion using a factor of 2.5, which is shown in Fig. \ref{fig:r_vs_s}. A significantly better (though still not perfect) correspondence is found using 2.5 as compared to 2.3. 

We adopt a factor of 2.5 for all fluxes in this work. The remaining disparity may be due to systematic differences in the ablation behaviour of different populations of meteoroids. Another potential source of error might be the power law assumption. For example, using a numerical model of meteoroid ejection from comet 1P/Halley, \cite{egal2020halleyids} predicted that Orionid meteoroids arriving at the Earth should have a mass distribution which follows a broken power law, notably with several sharp slope changes around 1~mm sizes.

After the fluxes in every time bin have been normalized to a single limiting magnitude of choice, an equivalent limiting mass can be computed. The magnitude to mass conversion used here is found in \citet{Peterson_Spacecraft_Book}. This relationship employs a speed-dependent luminous efficiency from \citet{verniani1965}, which \citet{jacchia1967} applied to Super-Schmidt meteors:

\begin{align}
    2.25 \log_{10}(m_{\mathrm{lim}}) = -8.75 \log_{10}\left({v_0}\right) - M_{\mathrm{lim}} +  11.59  \,,
\end{align}

\noindent where $m_{\mathrm{lim}}$ is the limiting mass in grams, $v_0$ is the initial in-atmosphere velocity in km/s, and $M_{\mathrm{lim}}$ is the peak meteor limiting magnitude in the panchromatic bandpass. As an example, for the Quadrantids with a velocity of 41 km/s the limiting mass at +6.5 magnitude is $9 \times 10^{-5}$ grams.

We estimate the errors in the computed flux using a classical assumption that the meteor events are distributed according to a Poisson distribution \citep{vida2020new}. A confidence interval for the flux based on the number of observed events can be computed using the ``exact'' method \citep{ulm1990simple}: $[\mathrm{Pr}(\alpha/2, 2N), \mathrm{Pr}(1-\alpha/2, 2(N + 1))]$, where $\mathrm{Pr}$ is the percent point function of the $\chi^2$ distribution, $N$ is the number of meteors, and $\alpha$ is the confidence interval ($\alpha=0.05$ for the 95\% confidence interval). The resulting range in the number of meteors expected based on Poisson statistics yields a flux confidence interval via Eq. \ref{eq:flux}, keeping the effective collection area fixed.

The zenithal hourly rate (ZHR) can be computed from the flux following \citet{koschack1990determination2}:

\begin{align}
    \mathrm{ZHR} = \frac{ 37200 F(6.5^\mathrm{M}) }{ (13.1 r - 16.5) (r - 1.3)^{0.748} }  
\end{align}

\noindent where $F(6.5^\mathrm{M})$ is the flux to the limiting meteor magnitude of $+6.5^\mathrm{M}$ in meteoroids~km$^{-2}$~h$^{-1}$.

\begin{figure*}
	\begin{center}
		\includegraphics[width=\linewidth]{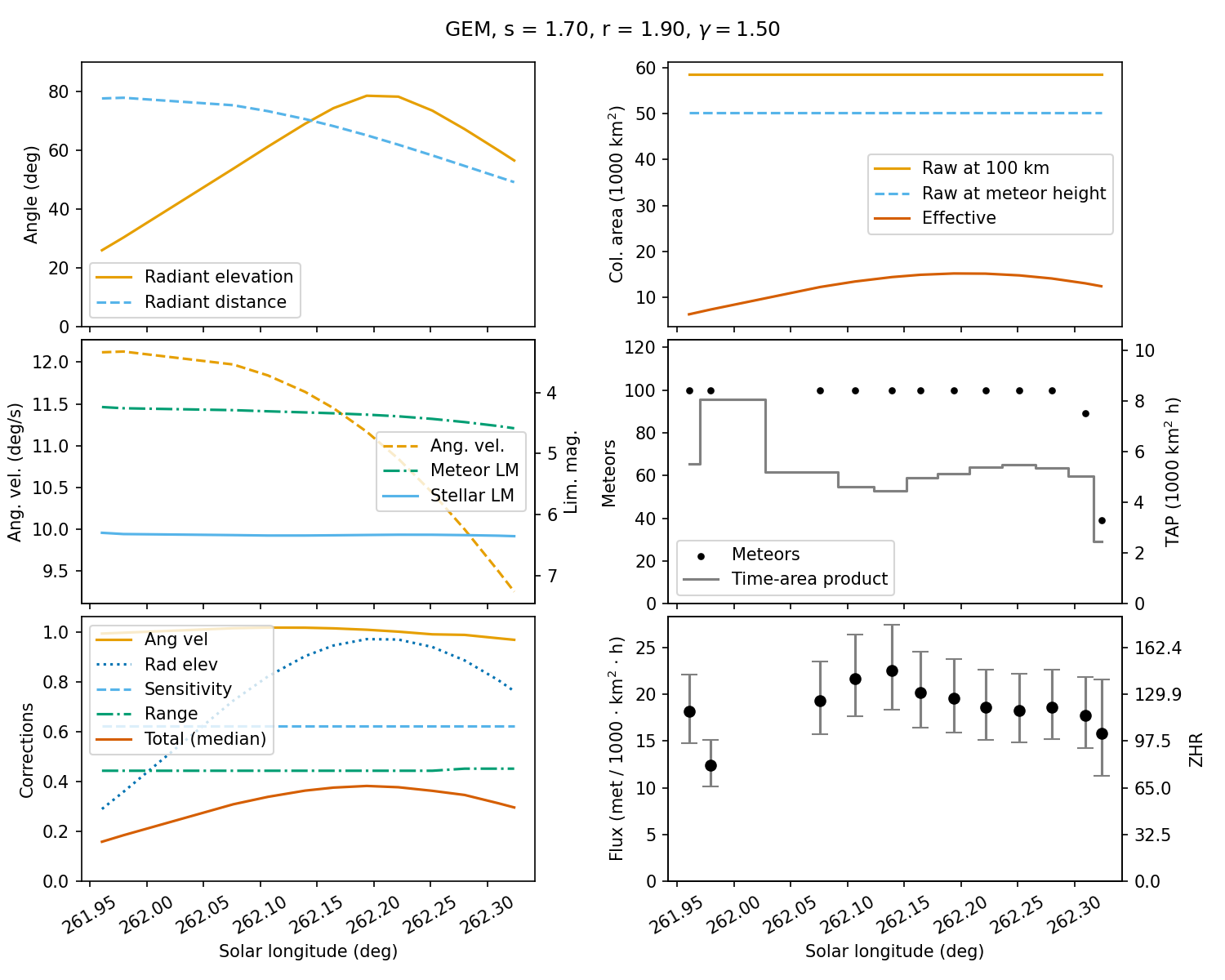}
	\end{center}
	\caption{An example showing the single-station flux of the 2020 Gemind peak observed by station HR000K. Individual computation steps and corrections are shown. The corrections shown are the median across the field of view. A fixed number of 100 meteors per bin is used, and the flux is computed to a limiting mass of $2\times10^{-4}$ grams (+6.5$^{\mathrm{M}}$).}
	\label{fig:flux_single_station}
\end{figure*}

Fig. \ref{fig:flux_single_station} is an example of the single-station flux output the software generated for the Geminids during the peak night of the shower. These reports are automatically generated for all showers relevant for spacecraft meteoroid impact risk assessment \citep{moorhead2019meteor}.

\subsubsection{Combining Single-station Measurement}

\begin{figure*}
	\begin{center}
		\includegraphics[width=\linewidth]{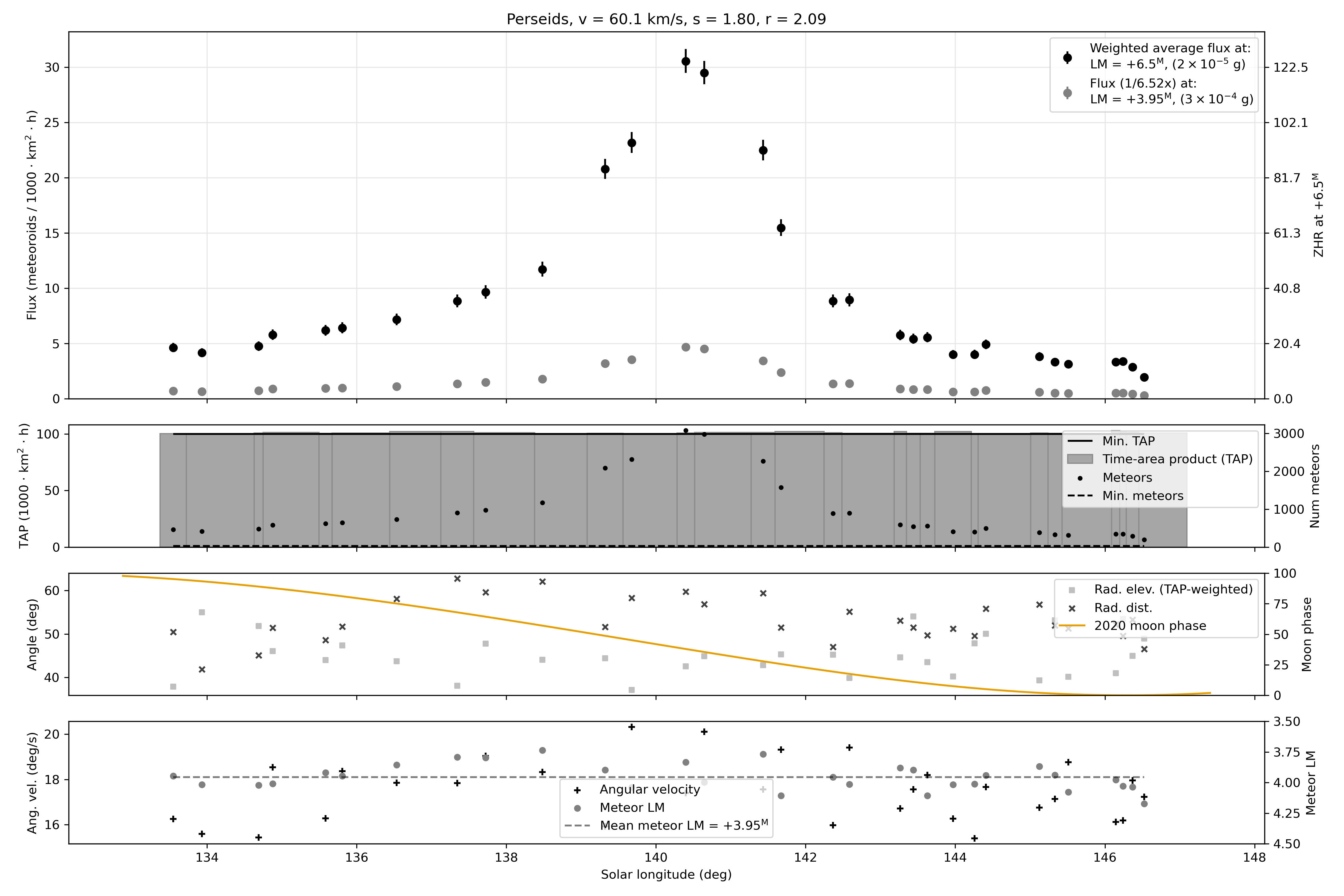}
	\end{center}
	\caption{Flux measurements of the 2020 Perseids using all cameras reporting data for the shower in the GMN (a total of 130 cameras in this example). A minimum TAP was set to 100,000~km$^2$~h to dynamically define the bin lengths. Error bars for some points in the wings are smaller than plot markers. Individual insets show mean TAP-weighted shower properties and limiting magnitude. Note that for this bin size and number of cameras each equivalent ''global'' measurement has between hundreds to thousands of meteors per data point. The top inset shows the mid-bin flux and confidence interval for a reference mass of $2 \times 10^{-5}$ g (black dots) while the grey symbols represent the flux to a limiting mass of $3 \times 10^{-4}$ g, which was derived from the average effective meteor limiting sensitivity of the aggregated network data ($+3.95^{\mathrm{M}}$). The former is scaled from the latter using the mass index of 1.8 as shown in the figure title. The second inset from top shows the time-area product (TAP) for the network as a function of solar longitude with the vertical grey lines denoting the time bins needed to achieve the constant TAP of 100,000~km$^2$~h. The black dots represent the total number of raw meteor detections by all cameras included in the flux calculation for a given bin. The third inset from the top shows the average radiant elevation across all cameras as well as the radiant distance from the center of the camera FOVs weighted by the TAP per bin. The phase of the moon is also shown for ease of interpretation, with 100 being full moon. Finally, the bottom inset shows the average limiting magnitude and average meteor shower angular velocity in the center of the field of view per bin, TAP weighted.}
	\label{fig:2020PER_flux}
\end{figure*}

To improve number statistics and produce robust and continuous flux measurements, we create a ``virtual" global instrument by combining single-station meteor counts and time-area product (TAPs) observations across the entire meteor network. The single-station flux code is run automatically on data from all available stations. To ensure that possible false positives (not real meteors) do not pollute the data set, all nights when more than 20 sporadic meteors per hour were detected are ignored. This value was found to be much higher than the number of real sporadic meteors ever observed in practice by any of the GMN cameras.

All observing periods are divided into fixed $\sim5$ minute bins (by solar longitude) which do not change across the years (if more than one year is being analysed). Meteor counts and TAPs are summed in each bin across all cameras. The implicit assumption is that all measurements have been scaled to the same limiting magnitude, which is achieved by scaling the TAP using Eq. \ref{eq:flux_scaling}. The meteor count remains as-is because it is used to compute the Poisson confidence interval.

Next, the 5 minute bins are combined to achieve a target minimum number of meteors and/or a minimum TAP per bin \citep{molau2014real, molau2020meteorflux}. This allows a user to choose between precision and time resolution. Fig. \ref{fig:2020PER_flux} shows the global instrument equivalent flux of the 2020 Perseids obtained by combining measurements across the whole network.

\subsubsection{Method Verification}

\begin{figure}
	\begin{center}
		\includegraphics[width=\linewidth]{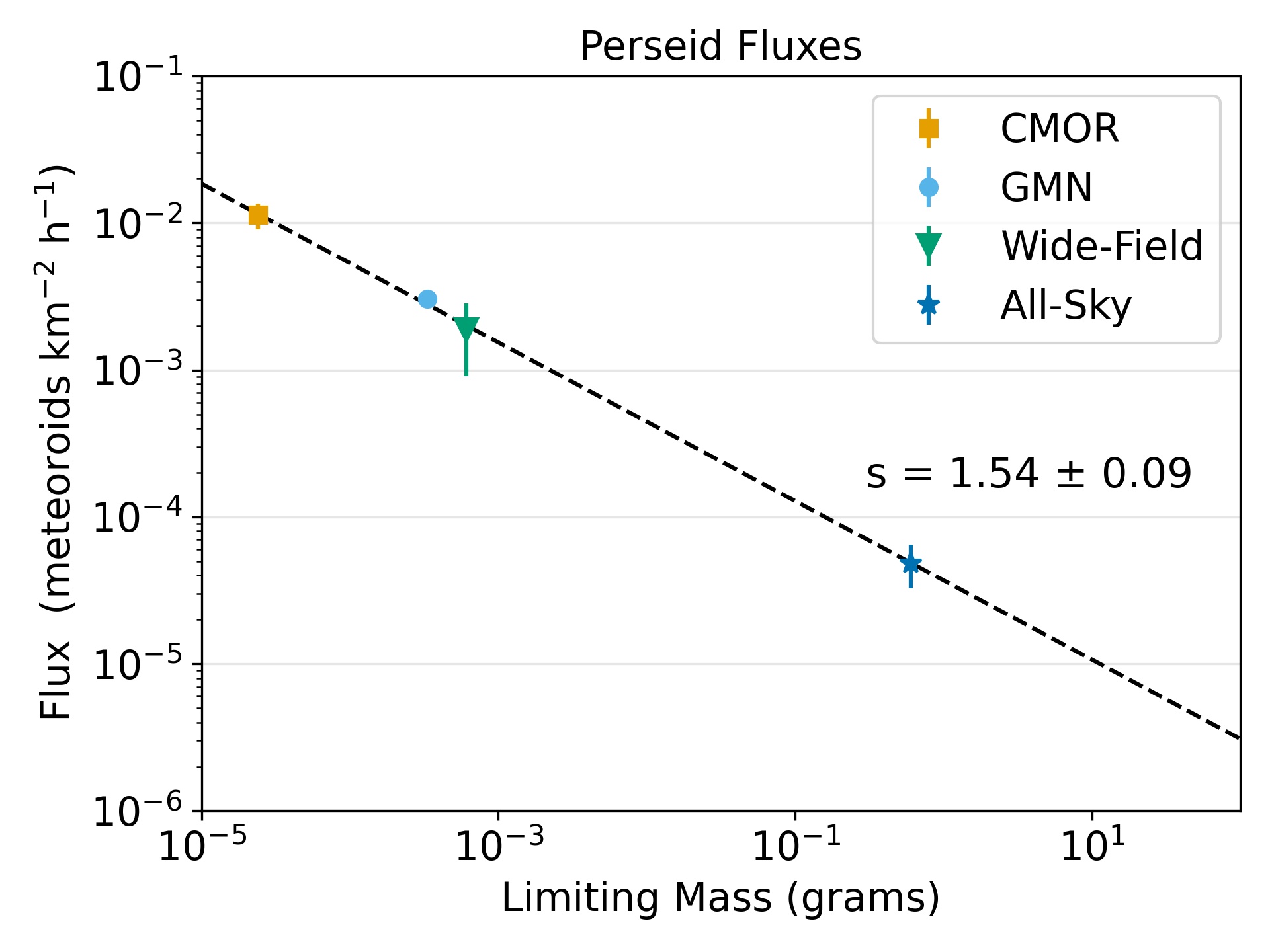}
	\end{center}
	\caption{Perseid fluxes from \citet{ehlert2020measuring} using measurements by the Canadian Orbit Meteor Radar (CMOR), NASA's Wide-Field meteor cameras and NASA's All-Sky meteor cameras. The GMN camera fluxes are also shown from this work. From these relative fluxes, a mass index of 1.54 $\pm$ 0.09 was measured. The GMN uncertainties are smaller than the symbol size. }
	\label{fig:PERmassIndex}
\end{figure}

To verify the code implementation, we compare the fluxes derived for two major showers with consistent year-to-year activity: the Perseids and the Geminids. A peak flux of $30.5^{+1.1}_{-1.1} \times 10^{-3}$ meteoroids~km$^{-2}$~h$^{-1}$ to a limiting mass and magnitude of $2.2 \times 10^{-5}$ g and $+6.5^{\mathrm{M}}$ ($4.68^{+.0.17}_{-.0.16} \times 10^{-3}$ meteoroids~km$^{-2}$~h$^{-1}$ at $3 \times 10^{-4}$ g, $+3.95^{\mathrm{M}}$) was measured for the Perseids in 2020 (ZHR = $124.7^{+4.5}_{-4.3}$, $s = 1.8$, $r = 2.1$) at solar longitude $140.4^{\circ} \pm 0.2^{\circ}$, as shown in Fig. \ref{fig:2020PER_flux}. This value compares favourably to the predicted flux at the given mass limit by \cite{ehlert2020measuring}, who based their model on independent measurements across 5 orders of magnitude in mass, as shown in Fig. \ref{fig:PERmassIndex}. This value is also consistent with both past visual and video flux measurements of the Perseids \citep{brown1996perseid, molau2019results}, as given in Table \ref{tab:perseid_flux}. The measured flux as a function of solar longitude is available in Supplementary Materials.

\begin{table*}
    \begin{tabular}{l | c | c | c | l | l | l | S[table-format=3]@{\,\( \pm \)\,} S[table-format=3] }
    \hline Study & Years & Component & Method & s & r & \multicolumn{1}{|p{3cm}|}{\centering Flux (met~km$^{-2}$~hr$^{-1}$)\\at $2.2\times10^{-5}$g ($+6.5^{\mathrm{M}}$) } & \multicolumn{2}{c}{ZHR}  \\
    \hline
    \citet{brown1996perseid}    & 1991 & Annual   & Visual & 1.80 & 2.10 & 0.024 $\pm$ 0.001$^{\dagger}$ &  97 &  2 \\
    \citet{brown1996perseid}    & 1992 & Annual   & Visual & 1.80 & 2.10 & 0.021 $\pm$ 0.009$^{\dagger}$ &  84 & 34 \\
    \citet{brown1996perseid}    & 1993 & Annual   & Visual & 1.80 & 2.10 & 0.022 $\pm$ 0.001$^{\dagger}$ &  86 &  2 \\
    \citet{brown1996perseid}    & 1994 & Annual   & Visual & 1.80 & 2.10 & 0.022 $\pm$ 0.001$^{\dagger}$ &  86 &  2 \\
    \citet{molau2019results}    & 2015 & Annual   & Video  & 1.80 & 2.20 & 0.038 $\pm$ 0.002*            & 124 &  5*\\
    \citet{molau2019results}    & 2018 & Annual   & Video  & 1.80 & 2.20 & 0.028 $\pm$ 0.001*            &  92 &  3*\\
    meteorflux.org              & 2020 & Annual   & Video  & 1.80 & 2.20 & 0.027 $\pm$ 0.001             &  86 &  5 \\
    VMDB                        & 2020 & Annual   & Visual & 2.00 & 2.50 & 0.039 $\pm$ 0.004$^{\dagger}$ &  77 &  7 \\
    Our study                   & 2020 & Annual   & Video  & 1.80 & 2.09 & 0.031 $\pm$ 0.001             & 125 &  5 \\
    \hline
    \citet{brown1996perseid}    & 1991 & Outburst & Visual & 1.80 & 2.10 & 0.071 $\pm$ 0.016$^{\dagger}$ & 284 &  63 \\
    \citet{jenniskens1998unusual}&1991 & Outburst & Photo  & 1.73 & 1.96 & 0.063 $\pm$ 0.014$^{\dagger}$ & 350 &  75 \\
    \citet{brown1996perseid}    & 1992 & Outburst & Visual & 1.80 & 2.10 & 0.055 $\pm$ 0.006$^{\dagger}$ & 220 &  22 \\
    \citet{jenniskens1998unusual}&1992 & Outburst & Photo  & 1.73 & 1.96 & 0.099 $\pm$ 0.027$^{\dagger}$ & 550 & 150 \\
    \citet{brown1996perseid}    & 1993 & Outburst & Visual & 1.80 & 2.10 & 0.066 $\pm$ 0.004$^{\dagger}$ & 264 &  17 \\
    \citet{jenniskens1998unusual}&1993 & Outburst & Photo  & 1.73 & 1.96 & 0.045 $\pm$ 0.006$^{\dagger}$ & 250 &  35 \\
    \citet{brown1996perseid}    & 1994 & Outburst & Visual & 1.80 & 2.10 & 0.060 $\pm$ 0.004$^{\dagger}$ & 238 &  17 \\
    \citet{jenniskens1998unusual}&1994 & Outburst & Photo  & 1.73 & 1.96 & 0.034 $\pm$ 0.005$^{\dagger}$ & 190 &  35 \\
    \citet{ehlert2020measuring} & 2016 & Outburst & Video  & 1.54 & 1.72 & 0.011 $\pm$ 0.003$^{\dagger}$ & 134 &  39 \\
\citet{miskotte2017magnificent} & 2016 & Outburst & Visual & 1.82 & 2.12 & 0.084 $\pm$ 0.004$^{\dagger}$ & 320 &  15* \\
    \citet{miskotte2021big}     & 2021 & Outburst & Radio  & 1.95 & 2.39 & 0.114 $\pm$ 0.003$^{\dagger}$ & 269 &   6* \\
    \citet{miskotte2021big}     & 2021 & Outburst & Visual & 2.10 & 2.76 & 0.136 $\pm$ 0.011$^{\dagger}$ & 195 &  16* \\
\citet{jenniskens2021perseid}   & 2021 & Outburst & Video  & 2.39 & 3.59 & 0.259 $\pm$ 0.030$^{\dagger}$ & 170 &  20* \\
    VMDB                        & 2021 & Outburst & Visual & 1.86 & 2.20 & 0.044 $\pm$ 0.003$^{\dagger}$ & 144 &  9 \\
    Our study                   & 2021 & Outburst & Video  & 1.80 & 2.09 & 0.068 $\pm$ 0.004             & 277 &  18 \\
    \end{tabular}
    \caption{Perseid flux measurements for the annual and the outburst component compared from this study to other previously published values. * - numbers taken from a plot, not given explicitly in text, $^{\dagger}$ - computed either from the flux or ZHR, whichever one was originally given in the paper.}
    \label{tab:perseid_flux}
\end{table*}

\begin{figure}
	\begin{center}
		\includegraphics[width=\linewidth]{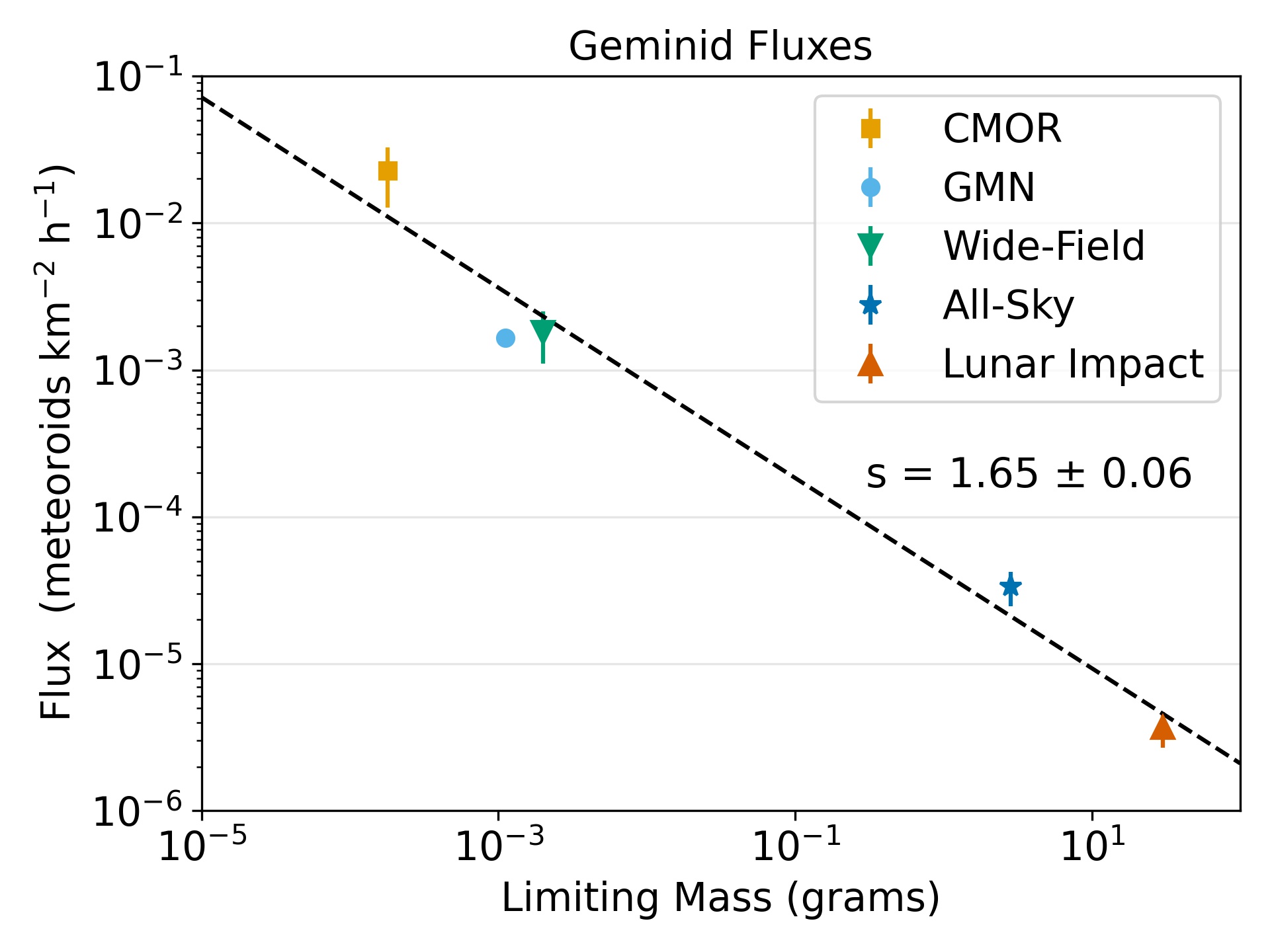}
	\end{center}
	\caption{Geminid fluxes from \citet{ehlert2020measuring} based on measurements by the Canadian Orbit Meteor Radar (CMOR), NASA's Wide-Field meteor cameras, NASA's All-Sky meteor cameras and NASA's Lunar Impact program. Also shown for comparison is the peak flux for the 2020-2021 Geminids from GMN cameras. With these fluxes a mass index of 1.65 $\pm$ 0.06 was measured. The GMN uncertainties are smaller than the symbol size. }
	\label{fig:GEMmassIndex}
\end{figure}

A combined peak flux of $20.04^{+.51}_{-.50} \times 10^{-3}$ meteoroids~km$^{-2}$~h$^{-1}$ at a limiting mass and magnitude of $2 \times 10^{-4}$ g and $+6.5^{\mathrm{M}}$ ($6.40^{+.16}_{-.16} \times 10^{-3}$ meteoroids~km$^{-2}$~h$^{-1}$ at $1 \times 10^{-3}$ g, $+4.73^{\mathrm{M}}$) was measured for the Geminids in 2020 and 2021 (ZHR = $128.2^{+3.3}_{-3.2}$, $s = 1.7$, $r = 1.9$) at solar longitude $261.98^{\circ} \pm 0.08^{\circ}$. The full flux profile is shown in Fig. \ref{fig:GEM_flux}, and the data are given in Supplementary Materials. These values are in accordance with recent visual and video flux measurements \citep{molau2015results, molau2016results, miskotte2019geminids}, as given in Table \ref{tab:geminid_flux}. \cite{ehlert2020measuring} measured their fluxes in 2015 the night after the peak, at solar longitude $\sim263.0$, and our measurement can be seen in Fig. \ref{fig:GEMmassIndex}. Our measurements lie below their curve by a factor of two, however the other data points across the large mass range show a similar scatter.

\begin{table*}
    \begin{tabular}{l | c | c | l | l | l | S[table-format=3.1]@{\,\( \pm \)\,} S[table-format=3.1] }
    \hline Study & Years & Method & s & r & \multicolumn{1}{|p{3cm}|}{\centering Flux (met~km$^{-2}$~hr$^{-1}$)\\at $2\times10^{-4}$g ($+6.5^{\mathrm{M}}$) } & \multicolumn{2}{c}{ZHR}  \\
    \hline
    \citet{molau2016results}    & 2012-2015 & Video  & 1.95 & 2.60 & 0.070 $\pm$ 0.005*            & 125   & 10*\\
    \citet{miskotte2019geminids}& 2018      & Visual & 1.69 & 1.80 & 0.014 $\pm$ 0.003$^{\dagger}$ & 121   & 22\\
    VMDB                        & 2020      & Visual & 1.87 & 2.40 & 0.058 $\pm$ 0.002$^{\dagger}$ & 134   &  5 \\
    meteorflux.org              & 2020      & Video  & 1.70 & 2.00 & 0.015 $\pm$ 0.001             &  76   &  3 \\
    VMDB                        & 2021      & Visual & 1.87 & 2.40 & 0.055 $\pm$ 0.002$^{\dagger}$ & 128   &  4 \\
    meteorflux.org              & 2021      & Video  & 1.70 & 2.00 & 0.018 $\pm$ 0.002             &  89   &  15 \\
    Our study                   & 2020-2021 & Video  & 1.70 & 1.91 & 0.020 $\pm$ 0.001             & 128.2 &  3.3 \\
    \end{tabular}
    \caption{Comparison of Geminid flux measurements from this study to other previously published values. * - numbers taken from a plot, not given explicitly in text, $^{\dagger}$ - computed either from the flux or ZHR, whichever one was originally given in the paper.}
    \label{tab:geminid_flux}
\end{table*}

\begin{figure*}
	\begin{center}
		\includegraphics[width=\linewidth]{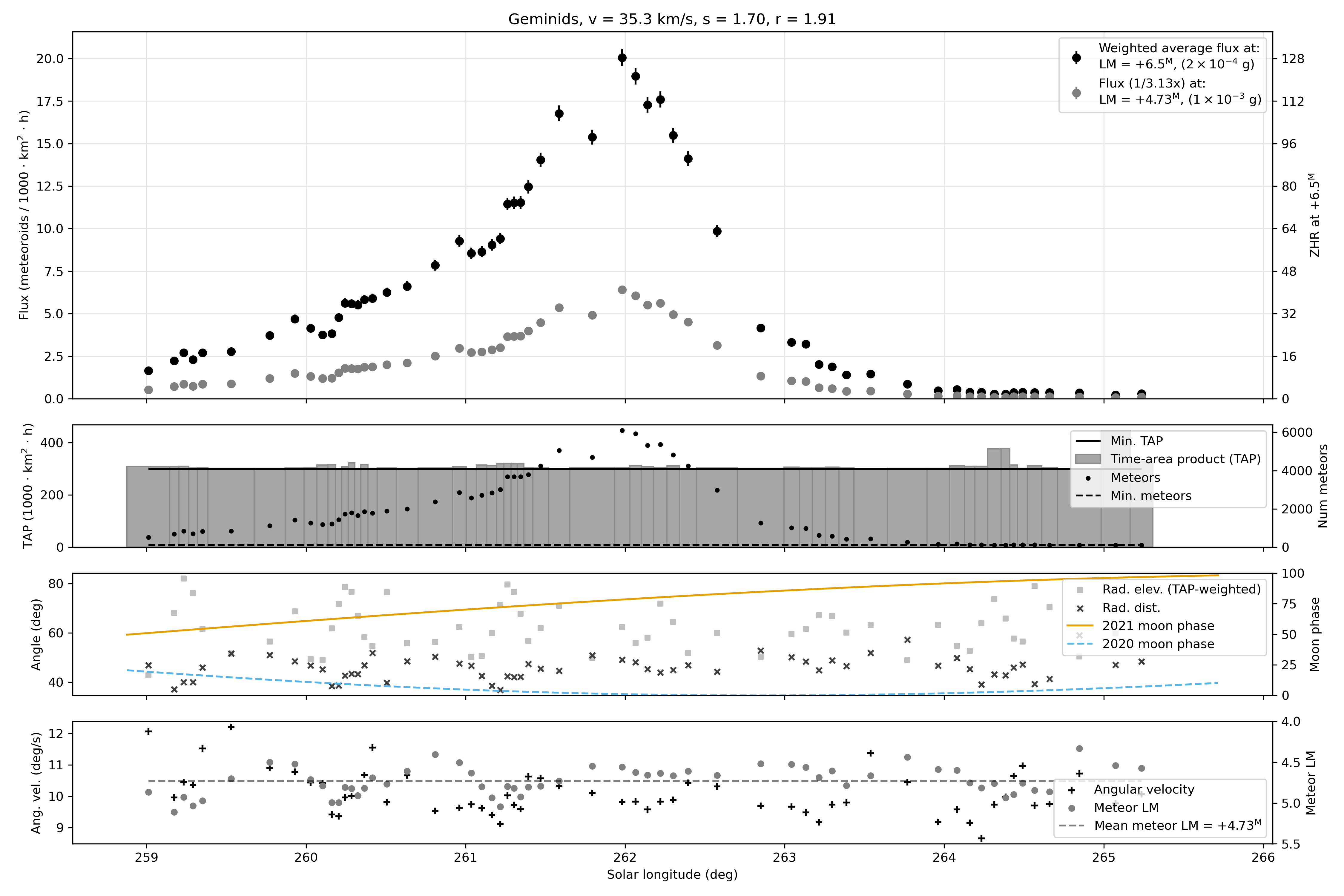}
	\end{center}
	\caption{Combined flux measurements of the 2020 and 2021 Geminids. A minimum TAP was set to 300,000~km$^2$~h and 100 meteors per bin. Error bars for some points in the wings are smaller than plot markers. The top inset shows the mid-bin flux and confidence interval for a reference mass and magnitude of $2 \times 10^{-4}$ g and $+6.5^{\mathrm{M}}$ (black dots), while the grey symbols represent the flux to a limiting mass of $1 \times 10^{-3}$ g, which is derived from the average effective meteor limiting sensitivity of the aggregated network data ($+4.73^{\mathrm{M}}$). The former is scaled from the latter using the mass index of 1.7, as shown in the figure title.}
	\label{fig:GEM_flux}
\end{figure*}

\section{Results} \label{sec:results}

Having verified our method with two showers showing consistent annual activity, we next present results for the more challenging situation of two recent showers/outbursts where fluxes varied on short timescales. Specifically we focus on the short-lived outburst reported for the 2021 Perseid return and examine the core of the Quadrantid meteor shower.

\subsection{2021 Perseid Outburst}

\begin{figure*}
	\begin{center}
		\includegraphics[width=\linewidth]{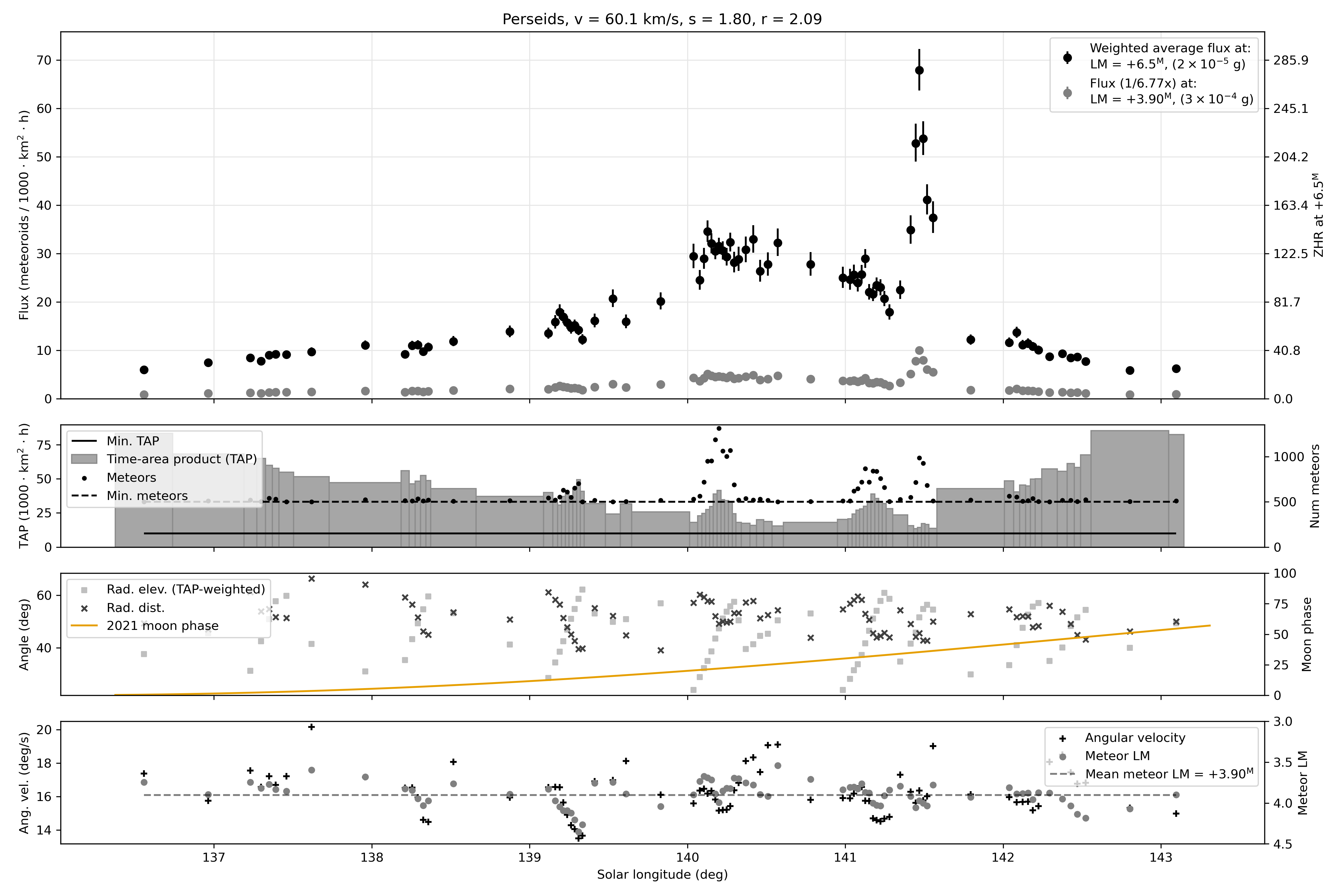}
	\end{center}
	\caption{Flux measurements of the 2021 Perseids showing the outburst. A minimum TAP was set to 10,000~km$^2$~h and a minimum of 500 meteors were taken per bin. The top inset shows the mid-bin flux and confidence interval for a reference mass and magnitude of $2 \times 10^{-5}$ g and $+6.5^{\mathrm{M}}$ (black dots), while the grey symbols represent the flux to a limiting mass of $3 \times 10^{-4}$ g, which is derived from the average effective meteor limiting sensitivity of the aggregated network data ($+3.90^{\mathrm{M}}$). The former is scaled from the latter using the mass index of 1.8 as shown in the figure title.}
	\label{fig:2021PER_outburst_flux}
\end{figure*}

Fig. \ref{fig:2021PER_outburst_flux} shows the flux of the 2021 Perseids recorded by the GMN. From this figure, the 2021 activity of the base Perseid component was consistent with the 2020 return, with a ZHR of $\sim120$. The peak flux of the outburst, however, was much higher having a flux of $67.9^{+4.4}_{-4.2} \times 10^{-3}$ meteoroids~km$^{-2}$~h$^{-1}$ to a limiting mass and magnitude of $2 \times 10^{-5}$ g and $+6.5^{\mathrm{M}}$ ($10.0^{+.7}_{-.6} \times 10^{-3}$ meteoroids~km$^{-2}$~h$^{-1}$ at $3 \times 10^{-4}$ g, $+3.90^{\mathrm{M}}$). 

The equivalent ZHR is $277^{+18}_{-17}$, in agreement with previous outbursts in 1991-1994 and 2016 \citep{brown1996perseid, jenniskens1998unusual, miskotte2017magnificent}, as given in Table \ref{tab:perseid_flux}. Our measurements are also in agreement with some radio and visual observations in 2021 \citep{miskotte2021big}, however other authors report a 25\% lower peak flux \citep[ZHR $\sim200$;][]{jenniskens2021perseid}. It is probable that due to the short duration of the outburst the chosen binning affects the peak values, and the uncertainty in the mass index which varies significantly in published work.

The peak time of the outburst occurred at solar longitude $141.470 \pm 0.024$, with a total duration of about two and a half hours. The end of the outburst and the subsequent $\sim10$ hours were not observed as the network had no coverage over the Pacific. The same mass index for the base component and the outburst of $s = 1.8$ was assumed. Due to the short duration of the outburst, it was only observed by 36 cameras in North America. The flux measurements are given in Supplementary Materials.

\subsection{Quadrantids 2020 - 2022}

\begin{figure*}
	\begin{center}
		\includegraphics[width=\linewidth]{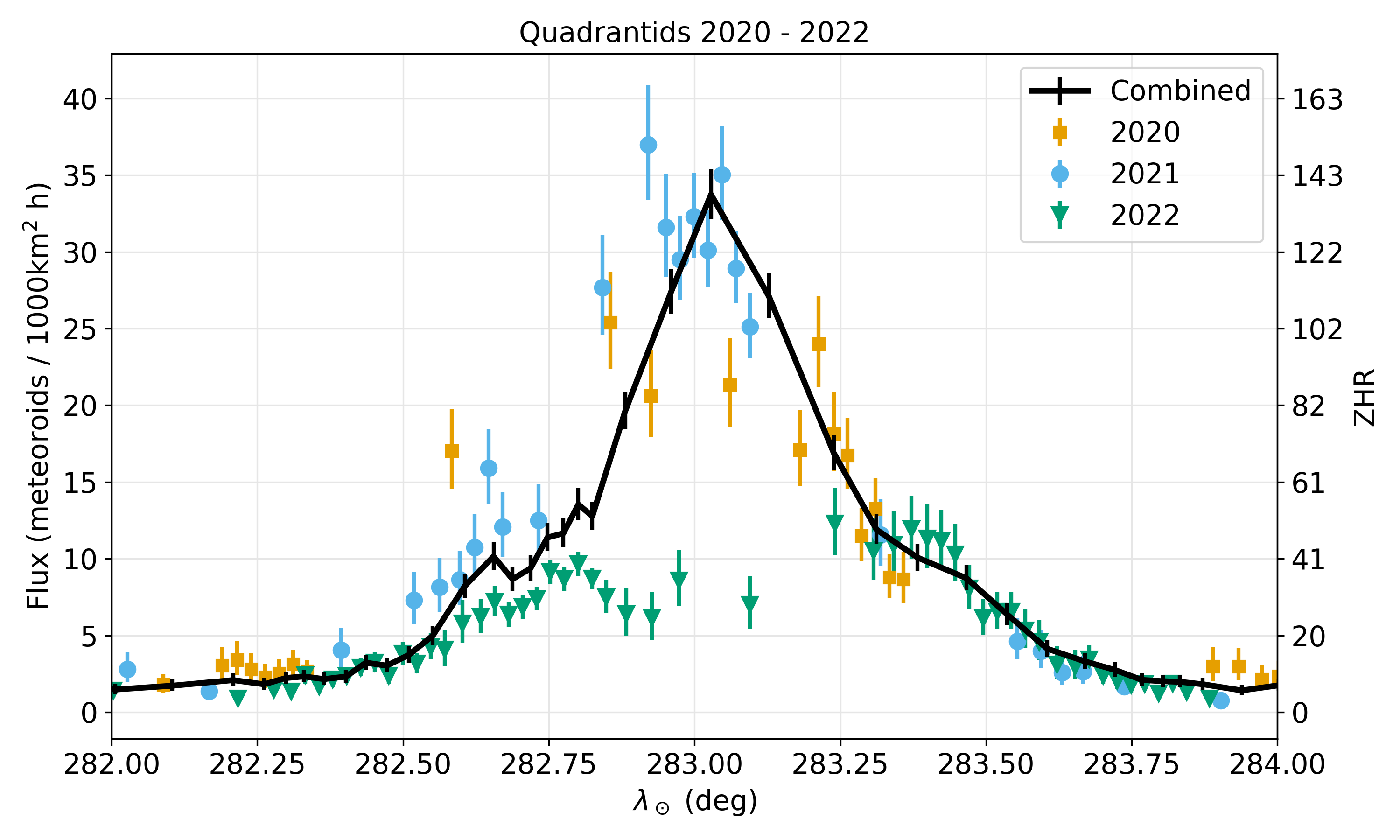}
	\end{center}
	\caption{Quadrantid fluxes to a limiting mass and magnitude of $9 \times 10^{-5}$ grams and $+6.5^{\mathrm{M}}$ found with the GMN camera network. The years 2020 - 2022 are shown individually as is their weighted average. $s = 1.8$ and $r = 2.09$ have been used to compute flux.}
	\label{fig:allQuadFlux}
\end{figure*}

Quadrantid fluxes from GMN cameras were combined between 2020-2022 to produce an average flux profile as presented in Fig. \ref{fig:allQuadFlux}, with individual years shown as well. These represent data from 69 cameras in 2020, 151 cameras in 2021, and 329 cameras in 2022. Taking a weighted average using all three years of data, the peak flux was found to be $33.7^{+1.6}_{-1.6} \times 10^{-3}$ meteoroids~km$^{-2}$~h$^{-1}$ to a limiting mass and magnitude of $9 \times 10^{-5}$ g and $+6.5^{\mathrm{M}}$ ($7.86^{+.38}_{-.37} \times 10^{-3}$ meteoroids~km$^{-2}$~h$^{-1}$ at $7 \times 10^{-4}$ grams, $+4.52^{\mathrm{M}}$). This equates to a ZHR of $137.8^{+6.7}_{-6.5}$. The peak occurred on average at $283.03^{\circ} \pm 0.1^{\circ}$ solar longitude. During 2022, the timing of the peak in flux occurred in regions which were mostly cloudy; hence the curve has few data points at the peak.

A mass index of $s = 1.8$ and a population index of $r = 2.1$ were used to calculate these Quadrantid fluxes. This value was chosen based on surveying 19 publications with calculated Quadrantid mass and population indices \citep{blaauw2011meteoroid}. This mass index has been used in other Quadrantid flux computations in the past, such as \citet{brown1998}. \citet{ehlert2020measuring} most recently constrained the Quadrantid mass index by measuring fluxes at three different mass limits and found $s = 1.83$.

Quadrantid fluxes have been measured for several past returns; results are summarized in Table \ref{tab:flux_compare}. \citet{brown1998} used radar to measure a flux of of 0.140 $\pm$ 0.010 meteoroids~km$^{-2}$~h$^{-1}$ at 283.08 $\pm$ $0.08^{\circ}$ solar longitude for the shower in 1997 to a limiting radio magnitude of $+7.7^{\mathrm{M}}$. This corresponded to a peak ZHR of almost 350. In another radar study, \citet{PooleEtAl_Quads} calculated the Quadrantid flux using 1964-1971 data, measuring a peak value of 0.056 meteoroids~km$^{-2}$~h$^{-1}$ down to $+6.5^{\mathrm{M}}$ (ZHR = $\sim490$).

In the optical mass range, the estimated equivalent rates are a factor of two smaller. Visual observations were analysed by \citet{rendtel1993} for the 1992 Quadrantids. They found a peak flux of 0.058 meteoroids~km$^{-2}$~h$^{-1}$ to $+6.5^{\mathrm{M}}$. This is equivalent to a peak ZHR of $\sim145$. They also measured a population index around the peak of $r = 2.1$, which corresponds to a mass index of 1.8 (using Eq. \ref{eq:mass_index}). Their peak occurred at $283.15^{\circ}$ solar longitude. Finally, \citet{ehlert2020measuring} measured the 2016 Quadrantid flux as $1.08 \times 10^{-5}$ meteoroids~km$^{-2}$~h$^{-1}$ to a limiting mass of 1.20 grams using all-sky cameras. This corresponds to a ZHR of 63 $\pm$ 33.

\citet{molau2016results} and \citet{molau2018results} presented Quadrantid fluxes from 2011 to 2018 using the IMO Video Network. While it is clear that the network did not catch the peak activity every year, during years where the peak was well covered, the ZHR ranges from 40 to 190, with 2015 showing the lowest peak flux, and 2014 showing the highest. The Quadrantid results found in the current study also display a variation in strength between 2020 and 2022. This may represent true yearly variation. The ascending and descending flux branches are very consistent, but the peak levels are variable, likely because the stream core is influenced by Jovian resonances \citep{MurrayEtAl_QUA}. This might also indicate an age difference between the two components, which is also evident from the shower radiant dispersion which is significantly higher during the time period away from the peak. The complete tables of measurements are given in Supplementary Materials.

For comparison, the Visual Meteor DataBase (VMDB)\footnote{International Meteor Organization: \url{www.imo.net}, accessed May 2, 2022.} reported peak ZHRs of 70.8 $\pm$ 2.6, 96.0 $\pm$ 14.8, and 44.1 $\pm$ 1.9 for 2020, 2021, and 2022, respectively. This follows the same trend as found by the GMN (peak ZHRs from 2020 to 2022 were $\sim100$, $\sim140$, and $\sim50$), with the highest flux observed in 2021, while 2022 was noticeably weaker. This underscores the wide variability in peak activity for the Quadrantids year to year which likely has an intrinsic component as well as a coverage aspect due to the short duration of the peak.

\begin{figure}
	\begin{center}
		\includegraphics[width=\linewidth]{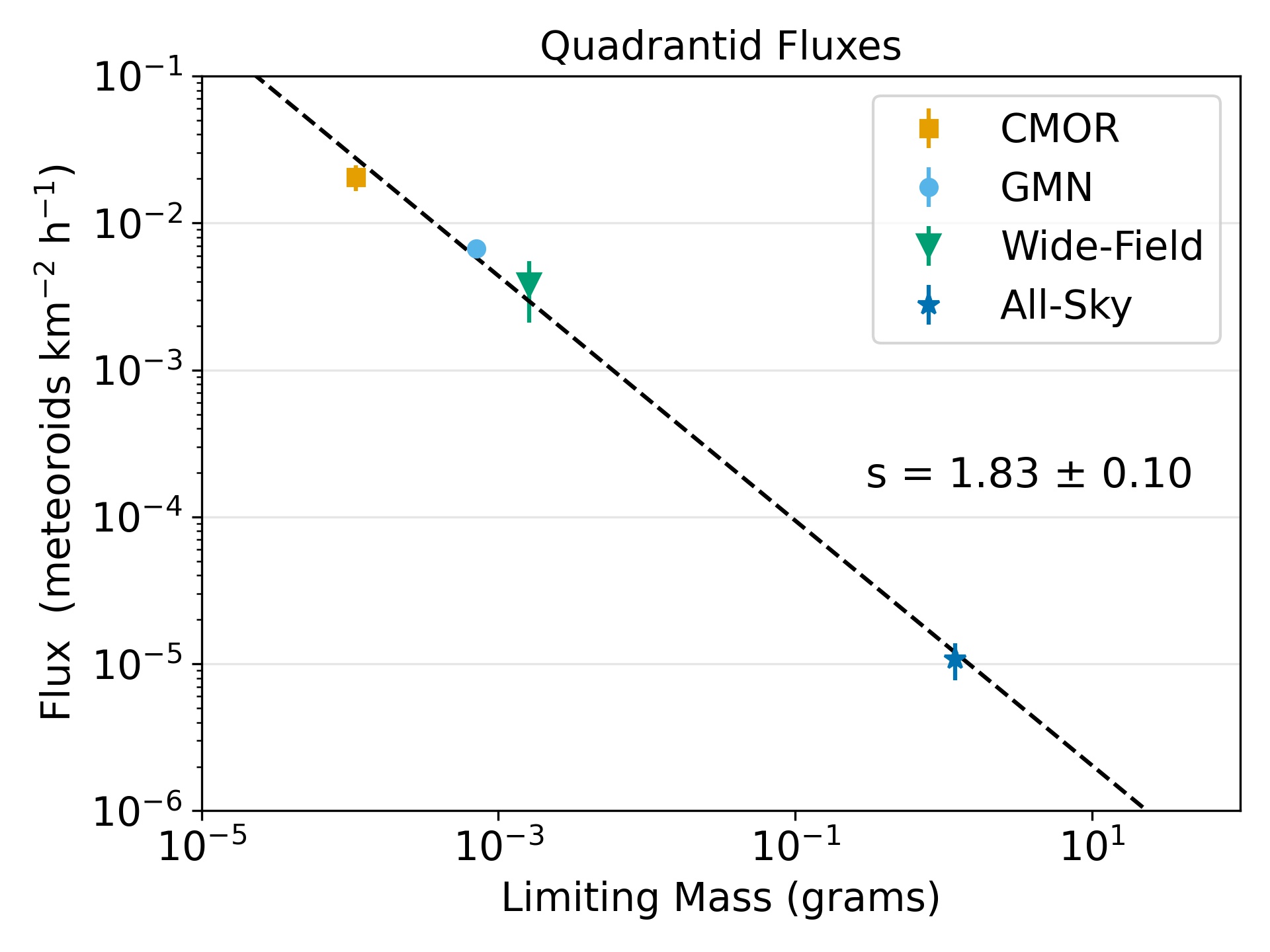}
	\end{center}
	\caption{Quadrantid fluxes from \citet{ehlert2020measuring} based on measurements by the Canadian Orbit Meteor Radar (CMOR), NASA's Wide-Field meteor cameras and NASA's All-Sky meteor cameras. Results for the average of the 2020-2022 returns from the GMN are also shown. From these fluxes as a function of mass, a mass index of 1.83 $\pm$ 0.10 was measured. The GMN fluxes have uncertainties smaller than the symbol size. }
	\label{fig:QUAmassIndex}
\end{figure}

Note that for each of these studies, the mass or population index plays a significant role in scaling the fluxes to arrive at an associated ZHR. Because of the sensitivity of fluxes to the mass index, it is difficult to compare studies that were calculated to different limiting masses. What can be helpful, however, is to plot them on a single plot, overlaying a constant mass index \citep{blaauw2017mass, ehlert2020measuring}, as shown in Fig. \ref{fig:QUAmassIndex}. The Canadian Meteor Orbit Radar (CMOR), Wide-field, and All-sky fluxes are extracted from \citet{ehlert2020measuring}. The GMN shower fluxes match well to these previous measurements. The mass index was determined using Monte Carlo linear regression simulations, and the uncertainty given is the standard deviations of the simulations.

\begin{table*}
    \begin{tabular}{l | c | c | l | l | l | S[table-format=3.1]@{\,\( \pm \)\,} S[table-format=2.1] }
    \hline Study & Years & Method & s & r & \multicolumn{1}{|p{3cm}|}{\centering Flux (met~km$^{-2}$~hr$^{-1}$)\\at $8.9\times10^{-5}$g ($+6.5^{\mathrm{M}}$) } & \multicolumn{2}{c}{ZHR}  \\
    \hline
    \citet{PooleEtAl_Quads}     & 1964-1971 & Radar  & 1.64 & 1.90 & 0.056*                        & 490 & ?$^{\dagger}$ \\
    \citet{rendtel1993}         & 1992      & Visual & 1.84 & 2.31 & 0.054 $\pm$ 0.001$^{\dagger}$ & 144.5  & 3.7 \\ 
    \citet{brown1998}           & 1997      & Radar  & 1.75 & 2.12 & 0.068 $\pm$ 0.008*            & 344    & 40$^{\dagger}$ \\
    \citet{brown1998}           & 1997      & Visual & 1.80 & 2.23 & 0.021 $\pm$ 0.015*            &  93    & 18* \\
    \citet{ehlert2020measuring} & 2016      & Video  & 1.83 & 2.30 & 0.017 $\pm$ 0.01$^{\dagger}$  &  63    & 33 \\
    meteorflux.org              & 2020-2022 & Video  & 1.80 & 2.20 & 0.024 $\pm$ 0.001             &  78    & 3 \\
    VMDB                        & 2020-2022 & Visual & 1.75 & 2.10 & 0.018 $\pm$ 0.002$^{\dagger}$ &  70    & 6.4 \\
    Our study                   & 2020-2022 & Video  & 1.80 & 2.09 & 0.030 $\pm$ 0.002             & 138    & 7  \\
    \end{tabular}
    \caption{Comparison of Quadrantid flux results from this study to other previously published values. * - numbers taken from a plot, not given explicitly in text, $^{\dagger}$ - computed either from the flux or ZHR, whichever one was originally given in the paper.}
    \label{tab:flux_compare}
\end{table*}

\section{Conclusions}

In this work we have presented details of a new optical meteor flux algorithm and applied it to data collected by the Global Meteor Network. A key feature of this approach is that it combines all cameras across a wide area into a single "virtual" global network improving number statistics and aggregating collection time-area products. 

The method was verified by comparing past studies to GMN-derived fluxes for two showers which show little year-to-year variation, namely the Geminids and the Perseids. From this analysis, we determined a peak flux of $30.5^{+1.1}_{-1.1} \times 10^{-3}$ meteoroids~km$^{-2}$~h$^{-1}$ to a limiting mass and magnitude of $2.2 \times 10^{-5}$ g and $+6.5^{\mathrm{M}}$ ($4.68^{+.0.17}_{-.0.16} \times 10^{-3}$ meteoroids~km$^{-2}$~h$^{-1}$ at $3 \times 10^{-4}$ g, $+3.95^{\mathrm{M}}$) for the annual component of the Perseids in 2020. This corresponds to a ZHR of $124.7^{+4.5}_{-4.3}$, assuming an $s = 1.8$, $r = 2.1$. The peak was found to be located at solar longitude $140.4^{\circ} \pm 0.2^{\circ}$. These values are in agreement with the general annual shower flux levels computed in other years from visual observations of the shower \citep{rendtel2014}. 

The flux measurements of the 2021 Perseid outburst using GMN data produced a peak ZHR of $277^{+18}_{-17}$ which also compares well to previous Perseid outbursts and is similar to the ZHR estimate of \citet{jenniskens2021perseid}.

From GMN data we measured a peak flux of $20.04^{+.51}_{-.50} \times 10^{-3}$ meteoroids~km$^{-2}$~h$^{-1}$ at a limiting mass and magnitude of $2 \times 10^{-4}$ g and $+6.5^{\mathrm{M}}$ ($6.40^{+.16}_{-.16} \times 10^{-3}$ meteoroids~km$^{-2}$~h$^{-1}$ at $1 \times 10^{-3}$ g, $+4.73^{\mathrm{M}}$) for the Geminids in 2020 and 2021 (ZHR = $128.2^{+3.3}_{-3.2}$, $s = 1.7$, $r = 1.9$) at solar longitude $261.98^{\circ} \pm 0.08^{\circ}$. This is also similar to the peak visual ZHR measured in recent years \citep{Ryabova2018}. A noticeable feature of our average profile is an apparent double peak in the flux near solar longitude $261.8^{\circ} \pm 0.2^{\circ}$, a feature predicted by several models of the stream \citep{Jones1982, Jones1985d, ryabova2007mathematical} and reported in visual data \citep{rendtel2014}.

Finally, we also determined the Quadrantid meteor shower flux between 2020-2022 from GMN data processed with the new flux methodology. The average peak flux was found to be $33.7^{+1.6}_{-1.6} \times 10^{-3}$ meteoroids~km$^{-2}$~h$^{-1}$ to a limiting mass and magnitude of $9 \times 10^{-5}$ g and $+6.5^{\mathrm{M}}$ ($7.86^{+.38}_{-.37} \times 10^{-3}$ meteoroids~km$^{-2}$~h$^{-1}$ at $7 \times 10^{-4}$ grams, $+4.52^{\mathrm{M}}$). This corresponds to a ZHR of $137.8^{+6.7}_{-6.5}$, again very similar to the peak ZHR reported from long-term visual records \citep{rendtel2014}.

\subsection{Note on Code Availability}

Implementation of all methods used in this work is published as open source on the following GitHub web pages:
\begin{itemize}
    \item RMS software: \\\url{https://github.com/CroatianMeteorNetwork/RMS}
    \item WesternMeteorPyLib: \\\url{https://github.com/wmpg/WesternMeteorPyLib}
\end{itemize}
Readers are encouraged to contact the lead author (DV) in the event they are not able to obtain the code on-line.

\subsection{Data Availability}

Per-station and summary flux measurements are available upon request. All flux data of the Perseids, Geminids, and the Quadrantids presented in this paper are given in supplementary materials.

\section{Acknowledgements}

Funding for this work was provided by the NASA Meteoroid Environment Office under cooperative agreement 80NSSC21M0073. The authors would like to thank Dr. Bill Cooke for providing insight and expertise that assisted the research and spurred good discussion. 

We would also like to thank the following GMN station operators whose cameras provided the data used in this work and contributors who made important code contributions (in alphabetical order): Richard Abraham, Victor Acciari, Rob Agar, David Akerman, Daknam Al-Ahmadi, Jamie Allen, Edison Álvarez José Felipe Pérezgómez, Alexandre Alves, Željko Andreić, Martyn Andrews, Miguel Angel Diaz, Enrique Arce, Georges Attard, Chris Baddiley, David Bailey, Erwin van Ballegoij, Roger Banks, Hamish Barker, Jean-Philippe Barrilliot, Richard Bassom, Ricky Bassom, Ehud Behar, Josip Belas, Alex Bell, Serge Bergeron, Denis Bergeron, Jorge Bermúdez Augusto Acosta, Steve Berry, Adrian Bigland, Chris Blake, Arie Blumenzweig, Ventsislav Bodakov, Claude Boivin, Robin Boivin, Bruno Bonicontro, Fabricio Borges, Ubiratan Borges, Dorian Božičević, David Brash, Ed Breuer, Martin Breukers, John W. Briggs, Peter G. Brown, Laurent Brunetto, Tim Burgess, Ludger Börgerding, Sylvain Cadieux, Peter Campbell-Burns, Andrew Campbell-Laing, Pablo Canedo, Seppe Canonaco, Jose Carballada, Steve Carter, David Castledine, Gilton Cavallini, Brian Chapman, Jason Charles, Matt Cheselka, Tim Claydon, Trevor Clifton, Manel Colldecarrera, Michael Cook, Bill Cooke, Christopher Coomber, Brendan Cooney, Edward Cooper, Jamie Cooper, Andrew Cooper, Marc Corretgé Gilart, Paul Cox, Llewellyn Cupido, Christopher Curtis, Ivica Ćiković, Dino Čaljkušić, Chris Dakin, Alfredo Dal’Ava Júnior, James Davenport, Richard Davis, Steve Dearden, Christophe Demeautis, Bart Dessoy, Paul Dickinson, Ivo Dijan, Pieter Dijkema, Tammo Jan Dijkema, Marcelo Domingues, Stacey Downton, Zoran Dragić, Iain Drea, Igor Duchaj, Jean-Paul Dumoulin, Garry Dymond, Jürgen Dörr, Robin Earl, Howard Edin, Ollie Eisman, Carl Elkins, Ian Graham Enting, Peter Eschman, Bob Evans, Andres Fernandez, Barry Findley, Rick Fischer, Richard Fleet, Jim Fordice, Jean Francois Larouche, Kyle Francis, Patrick Franks, Gustav Frisholm, Enrique Garcilazo Chávez, José García María, Mark Gatehouse, Ivan Gašparić, Chris George, Megan Gialluca, Kevin Gibbs-Wragge, Jason Gill, Philip Gladstone, Uwe Glässner, Hugo González, Nikola Gotovac, Neil Graham, Pete Graham, Colin Graham, Sam Green, Bob Greschke, Daniel J. Grinkevich, Larry Groom, Dominique Guiot, Tioga Gulon, Margareta Gumilar, Peter Gural, Nikolay Gusev, Kees Habraken, Alex Haislip, John Hale, Peter Hallett, Graeme Hanigan, Erwin Harkink, Ed Harman, Marián Harnádek, Ryan Harper, David Hatton, Tim Havens, Mark Haworth, Paul Haworth, Richard Hayler, Sam Hemmelgarn, Rick Hewett, Don Hladiuk, Alex Hodge, Simon Holbeche, Jeff Holmes, Nick Howarth, Matthew Howarth, Jeff Huddle, Bob Hufnagel, Roslina Hussain, Russell Jackson, Jean-Marie Jacquart, Jost Jahn, Phil James, Ron James Jr, Nick James, Rick James, Ilya Jankowsky, Alex Jeffery, Klaas Jobse, Richard Johnston, Dave Jones, Fernando Jordan, Vladimir Jovanović, Jocimar Justino de Souza, Javor Kac, Richard Kacerek, Milan Kalina, Jonathon Kambulow, Steve Kaufman, Paul Kavanagh, Alex Kichev, Harri Kiiskinen, Jean-Baptiste Kikwaya, Sebastian Klier, Dan Klinglesmith, John Kmetz, Zoran Knez, Korado Korlević, Stanislav Korotkiy, Danko Kočiš, Josip Krpan, Zbigniew Krzeminski, Patrik Kukić, Reinhard Kühn, Remi Lacasse, Gaétan Laflamme, Steve Lamb, Hervé Lamy, Jean Larouche Francois, Ian Lauwerys, David Leurquin, Gareth Lloyd, Rob Long De Corday, Eric Lopez, Jose Lopez Galindo, José Luis Martin, Juan Luis Muñoz, Pete Lynch, Frank Lyter, Anton Macan, Jonathan Mackey, John Maclean, Igor Macuka, Nawaz Mahomed, Simon Maidment, Mirjana Malarić, Nedeljko Mandić, Alain Marin, Bob Marshall, Colin Marshall, Andrei Marukhno, Keith Maslin, Nicola Masseroni, Bob Massey, Jacques Masson, Damir Matković, Filip Matković, Dougal Matthews, Michael Mazur, Sergio Mazzi, Stuart McAndrew, Alex McConahay, Robert McCoy, Charlie McCormack, Mark McIntyre, Peter Meadows, Edgar Merizio Mendes, Aleksandar Merlak, Filip Mezak, Pierre-Michael Micaletti, Greg Michael, Matej Mihelčić, Simon Minnican, Wullie Mitchell, Edson Morales Valencia, Nick Moskovitz, Dave Mowbray, Andrew Moyle, Gene Mroz, Muhammad Luqmanul Hakim Muharam, Brian Murphy, Carl Mustoe, Przemek Nagański, Jean-Louis Naudin, Damjan Nemarnik, Attila Nemes, Dave Newbury, Colin Nichols, Nick Norman, Philip Norton, Zoran Novak, Gareth Oakey, Washington Oliveira, Angélica Olmos López, Jamie Olver, Christine Ord, Nigel Owen, Michael O’Connell, Dylan O’Donnell, Thiago Paes, Carl Panter, Neil Papworth, Filip Parag, Gary Parker, Simon Parsons, Ian Pass, Igor Pavletić, Lovro Pavletić, Richard Payne, Pierre-Yves Pechart, William Perkin, Enrico Pettarin, Alan Pevec, Mark Phillips, Patrick Poitevin, Tim Polfliet, Pierre de Ponthière, Derek Poulton, Janusz Powazki, Aled Powell, Alex Pratt, Miguel Preciado, Chuck Pullen, Terry Pundiak, Lev Pustil’Nik, Dan Pye, Chris Ramsay, David Rankin, Steve Rau, Dustin Rego, Chris Reichelt, Danijel Reponj, Fernando Requena, Maciej Reszelsk, Ewan Richardson, Martin Richmond-Hardy, Mark Robbins, David Robinson, Martin Robinson, Heriton Rocha, Herve Roche, Adriana Roggemans, Paul Roggemans, Alex Roig, David Rollinson, Jim Rowe, Dmitrii Rychkov, Michel Saint-Laurent, Clive Sanders, Jason Sanders, Ivan Sardelić, Rob Saunders, John Savage, Lawrence Saville, Vasilii Savtchenko, William Schauff, Ansgar Schmidt, Jim Seargeant, Jay Shaffer, Steven Shanks, Mike Shaw, Angel Sierra, Ivo Silvestri, Ivica Skokić, Dave Smith, Tracey Snelus, Warley Souza, Mark Spink, Denis St-Gelais, James Stanley, Radim Stano, Laurie Stanton, Robert D. Steele, Yuri Stepanychev, Graham Stevens, Peter Stewart, William Stewart, Con Stoitsis, Andrea Storani, Andy Stott, David Strawford, Rajko Sušanj, Bela Szomi Kralj, Marko Šegon, Damir Šegon, Jeremy Taylor, Yakov Tchenak, Eric Toops, Torcuill Torrance, Steve Trone, Wenceslao Trujillo, John Tuckett, Myron Valenta, Jean Vallieres, Paraksh Vankawala, Neville Vann, Marco Verstraaten, Arie Verveer, Predrag Vukovic, Martin Walker, Aden Walker, Bill Wallace, John Waller, Jacques Walliang, Didier Walliang, Jacques Walliang, Christian Wanlin, Tom Warner, Neil Waters, Steve Welch, Alexander Wiedekind-Klein, John Wildridge, Ian Williams, Guy Williamson, Urs Wirthmueller, Bill Witte, Martin Woodward, Jonathan Wyatt, Anton Yanishevskiy, Penko Yordanov, Stephane Zanoni, and Dario Zubović.




\bibliographystyle{mnras}
\bibliography{bibliography} 

\begin{thebibliography}{}
\makeatletter
\relax
\def\mn@urlcharsother{\let\do\@makeother \do\$\do\&\do\#\do\^\do\_\do\%\do\~}
\def\mn@doi{\begingroup\mn@urlcharsother \@ifnextchar [ {\mn@doi@}
  {\mn@doi@[]}}
\def\mn@doi@[#1]#2{\def\@tempa{#1}\ifx\@tempa\@empty \href
  {http://dx.doi.org/#2} {doi:#2}\else \href {http://dx.doi.org/#2} {#1}\fi
  \endgroup}
\def\mn@eprint#1#2{\mn@eprint@#1:#2::\@nil}
\def\mn@eprint@arXiv#1{\href {http://arxiv.org/abs/#1} {{\tt arXiv:#1}}}
\def\mn@eprint@dblp#1{\href {http://dblp.uni-trier.de/rec/bibtex/#1.xml}
  {dblp:#1}}
\def\mn@eprint@#1:#2:#3:#4\@nil{\def\@tempa {#1}\def\@tempb {#2}\def\@tempc
  {#3}\ifx \@tempc \@empty \let \@tempc \@tempb \let \@tempb \@tempa \fi \ifx
  \@tempb \@empty \def\@tempb {arXiv}\fi \@ifundefined
  {mn@eprint@\@tempb}{\@tempb:\@tempc}{\expandafter \expandafter \csname
  mn@eprint@\@tempb\endcsname \expandafter{\@tempc}}}

\bibitem[\protect\citeauthoryear{Abedin, Spurn{\`y}, Wiegert, Pokorn{\`y},
  Borovi{\v{c}}ka  \& Brown}{Abedin et~al.}{2015}]{abedin2015}
Abedin A.,  Spurn{\`y} P.,  Wiegert P.,  Pokorn{\`y} P.,  Borovi{\v{c}}ka J.,
  Brown P.,  2015, Icarus, 261, 100

\bibitem[\protect\citeauthoryear{Abedin, Wiegert, Janches, Pokorn{\'{y}}, Brown
   \& Hormaechea}{Abedin et~al.}{2018}]{abedin2018}
Abedin A.,  Wiegert P.,  Janches D.,  Pokorn{\'{y}} P.,  Brown P.~G.,
  Hormaechea J.~L.,  2018, \mn@doi [Icarus] {10.1016/j.icarus.2017.07.015},
  300, 360

\bibitem[\protect\citeauthoryear{Beech \& Brown}{Beech \&
  Brown}{1993}]{beech1993impact}
Beech M.,  Brown P.,  1993, Monthly Notices of the Royal Astronomical Society,
  262, L35

\bibitem[\protect\citeauthoryear{Belkovich \& Tohktasev}{Belkovich \&
  Tohktasev}{1974}]{belkovich1974determination}
Belkovich O.,  Tohktasev V.,  1974, Bulletin of the Astronomical Institutes of
  Czechoslovakia, 25, 370

\bibitem[\protect\citeauthoryear{Blaauw}{Blaauw}{2017}]{blaauw2017mass}
Blaauw R.,  2017, Planetary and Space Science, 143, 83

\bibitem[\protect\citeauthoryear{Blaauw, Campbell-Brown  \& Weryk}{Blaauw
  et~al.}{2011}]{blaauw2011meteoroid}
Blaauw R.,  Campbell-Brown M.,   Weryk R.,  2011, Monthly Notices of the Royal
  Astronomical Society, 414, 3322

\bibitem[\protect\citeauthoryear{Blaauw, Campbell-Brown  \& Kingery}{Blaauw
  et~al.}{2016}]{blaauw2016optical}
Blaauw R.,  Campbell-Brown M.,   Kingery A.,  2016, Monthly Notices of the
  Royal Astronomical Society, 463, 441

\bibitem[\protect\citeauthoryear{Brown \& Jones}{Brown \&
  Jones}{1998}]{brown1998simulation}
Brown P.,  Jones J.,  1998, Icarus, 133, 36

\bibitem[\protect\citeauthoryear{Brown \& Rendtel}{Brown \&
  Rendtel}{1996}]{brown1996perseid}
Brown P.,  Rendtel J.,  1996, Icarus, 124, 414

\bibitem[\protect\citeauthoryear{Brown, Hocking, Jones  \& Rendtel}{Brown
  et~al.}{1998}]{brown1998}
Brown P.,  Hocking W.,  Jones J.,   Rendtel J.,  1998, Monthly Notices of the
  Royal Astronomical Society, 295, 847

\bibitem[\protect\citeauthoryear{Brown, Campbell, Hawkes, Theijsmeijer  \&
  Jones}{Brown et~al.}{2002}]{Brown2002b}
Brown P.,  Campbell M.,  Hawkes R.,  Theijsmeijer C.,   Jones J.,  2002,
  Planetary and Space Science, 50, 45

\bibitem[\protect\citeauthoryear{Campbell-Brown \& Jones}{Campbell-Brown \&
  Jones}{2006}]{campbell2006annual}
Campbell-Brown M.,  Jones J.,  2006, Monthly Notices of the Royal Astronomical
  Society, 367, 709

\bibitem[\protect\citeauthoryear{Caswell, McBride  \& Taylor}{Caswell
  et~al.}{1995}]{caswell1995olympus}
Caswell R.~D.,  McBride N.,   Taylor A.,  1995, International Journal of Impact
  Engineering, 17, 139

\bibitem[\protect\citeauthoryear{Ceplecha, Borovi{\v{c}}ka, Elford, ReVelle,
  Hawkes, Porub{\v{c}}an  \& {\v{S}}imek}{Ceplecha
  et~al.}{1998}]{ceplecha1998meteor}
Ceplecha Z.,  Borovi{\v{c}}ka J.,  Elford W.~G.,  ReVelle D.~O.,  Hawkes R.~L.,
   Porub{\v{c}}an V.,   {\v{S}}imek M.,  1998, Space Science Reviews, 84, 327

\bibitem[\protect\citeauthoryear{Duffy, Hawkes  \& Jones}{Duffy
  et~al.}{1987}]{Duffy1987}
Duffy A.,  Hawkes R.,   Jones J.,  1987, Monthly Notices of the Royal
  Astronomical Society, 228, 55

\bibitem[\protect\citeauthoryear{Egal, Wiegert, Brown, Campbell-Brown  \&
  Vida}{Egal et~al.}{2020}]{egal2020halleyids}
Egal A.,  Wiegert P.,  Brown P.~G.,  Campbell-Brown M.,   Vida D.,  2020,
  Astronomy \& Astrophysics, 642, A120

\bibitem[\protect\citeauthoryear{Ehlert \& Erskine}{Ehlert \&
  Erskine}{2020}]{ehlert2020measuring}
Ehlert S.,  Erskine R.~B.,  2020, Planetary and Space Science, p. 104938

\bibitem[\protect\citeauthoryear{Fulle et~al.,}{Fulle
  et~al.}{2016}]{fulle2016evolution}
Fulle M.,  et~al., 2016, The Astrophysical Journal, 821, 19

\bibitem[\protect\citeauthoryear{{Gaia Collaboration} et~al.,}{{Gaia
  Collaboration} et~al.}{2018}]{gaia2018dr2}
{Gaia Collaboration} et~al., 2018, \mn@doi [A\&A]
  {10.1051/0004-6361/201833051}, 616, A1

\bibitem[\protect\citeauthoryear{Galligan \& Baggaley}{Galligan \&
  Baggaley}{2004}]{galligan2004orbital}
Galligan D.,  Baggaley W.,  2004, Monthly Notices of the Royal Astronomical
  Society, 353, 422

\bibitem[\protect\citeauthoryear{Gr{\"{u}}n, Baguhl, Svedhem  \&
  Zook}{Gr{\"{u}}n et~al.}{2002}]{Grun2002}
Gr{\"{u}}n E.,  Baguhl M.,  Svedhem H.,   Zook H.,  2002, in Grun E.,
  Gustafson B.,  Dermott S.,   Fechtig H.,  eds, , Interplanetary Dust.
Springer Berlin Heidelberg, pp 295--346

\bibitem[\protect\citeauthoryear{Gural}{Gural}{2002}]{gural2002meteor}
Gural P.~S.,  2002, in Proceedings of the International Meteor Conference, 20th
  IMC, Cerkno, Slovenia, 2001. pp 29--35

\bibitem[\protect\citeauthoryear{{Gural}}{{Gural}}{2011}]{gural2011california}
{Gural} P.~S.,  2011, in Proceedings of the International Meteor Conference,
  29th IMC, Armagh, Northern Ireland, 2010. pp 28--31

\bibitem[\protect\citeauthoryear{Hawkes}{Hawkes}{2002}]{Hawkes2002}
Hawkes R.,  2002, in Murad E.,  Williams I.,  eds, , Meteors in the Earth's
  Atmosphere.
Cambridge Univ Press, Chapt.~5, pp 97--122, \url
  {http://adsabs.harvard.edu/abs/2002mea..book...97H}

\bibitem[\protect\citeauthoryear{Hughes \& McBride}{Hughes \&
  McBride}{1989}]{hughes1989mass}
Hughes D.~W.,  McBride N.,  1989, Monthly Notices of the Royal Astronomical
  Society, 240, 73

\bibitem[\protect\citeauthoryear{{Jacchia}, {Verniani}  \& {Briggs}}{{Jacchia}
  et~al.}{1967}]{jacchia1967}
{Jacchia} L.~G.,  {Verniani} F.,   {Briggs} R.~E.,  1967, Smithsonian
  Contributions to Astrophysics, \href
  {https://ui.adsabs.harvard.edu/abs/1967SCoA...11....1J} {11, 1}

\bibitem[\protect\citeauthoryear{Jenniskens \& Miskotte}{Jenniskens \&
  Miskotte}{2021}]{jenniskens2021perseid}
Jenniskens P.,  Miskotte K.,  2021, eMeteorNews, 6, 460

\bibitem[\protect\citeauthoryear{Jenniskens et~al.,}{Jenniskens
  et~al.}{1998}]{jenniskens1998unusual}
Jenniskens P.,  et~al., 1998, Monthly Notices of the Royal Astronomical
  Society, 301, 941

\bibitem[\protect\citeauthoryear{Jenniskens et~al.,}{Jenniskens
  et~al.}{2016}]{jenniskens2016cams}
Jenniskens P.,  et~al., 2016, Icarus, 266, 384

\bibitem[\protect\citeauthoryear{Jewitt \& Li}{Jewitt \&
  Li}{2010}]{jewitt2010activity}
Jewitt D.,  Li J.,  2010, The Astronomical Journal, 140, 1519

\bibitem[\protect\citeauthoryear{Jones}{Jones}{1982}]{Jones1982}
Jones J.,  1982, \mn@doi [Monthly Notices of the Royal Astronomical Society]
  {10.1093/mnras/198.1.23}, 198, 23

\bibitem[\protect\citeauthoryear{Jones}{Jones}{1985}]{Jones1985d}
Jones J.,  1985, Monthly Notices of the Royal Astronomical Society, 217, 523

\bibitem[\protect\citeauthoryear{Kaiser}{Kaiser}{1960}]{Kaiser1960}
Kaiser T.,  1960, Monthly Notices of the Royal Astronomical Society, 121, 284

\bibitem[\protect\citeauthoryear{Koschack \& Rendtel}{Koschack \&
  Rendtel}{1990a}]{koschack1990determination1}
Koschack R.,  Rendtel J.,  1990a, WGN, Journal of the International Meteor
  Organization, 18, 44

\bibitem[\protect\citeauthoryear{Koschack \& Rendtel}{Koschack \&
  Rendtel}{1990b}]{koschack1990determination2}
Koschack R.,  Rendtel J.,  1990b, WGN, Journal of the International Meteor
  Organization, 18, 119

\bibitem[\protect\citeauthoryear{Koten, Rendtel, Shrben\'{y}, Gural, Borovicka
  \& Kozak}{Koten et~al.}{2019}]{koten2019meteors}
Koten P.,  Rendtel J.,  Shrben\'{y} L.,  Gural P.,  Borovicka J.,   Kozak P.,
  2019, Meteoroids: Sources of Meteors on Earth and Beyond, pp 90--115

\bibitem[\protect\citeauthoryear{{Miskotte}}{{Miskotte}}{2019}]{miskotte2019geminids}
{Miskotte} K.,  2019, eMeteorNews, \href
  {https://ui.adsabs.harvard.edu/abs/2019eMetN...4..207M} {4, 207}

\bibitem[\protect\citeauthoryear{{Miskotte} \& {Vandeputte}}{{Miskotte} \&
  {Vandeputte}}{2017}]{miskotte2017magnificent}
{Miskotte} K.,  {Vandeputte} M.,  2017, eMeteorNews, \href
  {https://ui.adsabs.harvard.edu/abs/2017eMetN...2...61M} {2, 61}

\bibitem[\protect\citeauthoryear{{Miskotte}, {Sugimoto}  \&
  {Martin}}{{Miskotte} et~al.}{2021}]{miskotte2021big}
{Miskotte} K.,  {Sugimoto} H.,   {Martin} P.,  2021, eMeteorNews, \href
  {https://ui.adsabs.harvard.edu/abs/2021eMetN...6..517M} {6, 517}

\bibitem[\protect\citeauthoryear{Molau}{Molau}{2020}]{molau2020meteorflux}
Molau S.,  2020, in Proceedings of the International Meteor Conference,
  Virtual. pp 57--59

\bibitem[\protect\citeauthoryear{Molau \& Barentsen}{Molau \&
  Barentsen}{2013}]{molau2013meteoroid}
Molau S.,  Barentsen G.,  2013, in Proceedings of the International Meteor
  Conference, La Palma, Canary Islands, Spain, 2012. pp 11--17

\bibitem[\protect\citeauthoryear{Molau \& Barentsen}{Molau \&
  Barentsen}{2014}]{molau2014real}
Molau S.,  Barentsen G.,  2014, Earth, Moon, and Planets, 112, 1

\bibitem[\protect\citeauthoryear{Molau, Barentsen  \& Crivello}{Molau
  et~al.}{2014}]{molau2014obtaining}
Molau S.,  Barentsen G.,   Crivello S.,  2014, in Proceedings of the
  International Meteor Conference, Giron, France. pp 18--21

\bibitem[\protect\citeauthoryear{{Molau} et~al.,}{{Molau}
  et~al.}{2015}]{molau2015results}
{Molau} S.,  et~al., 2015, WGN, Journal of the International Meteor
  Organization, \href {https://ui.adsabs.harvard.edu/abs/2015JIMO...43...85M}
  {43, 85}

\bibitem[\protect\citeauthoryear{{Molau}, {Crivello}, {Goncalves}, {Saraiva},
  {Stomeo}  \& {Kac}}{{Molau} et~al.}{2016}]{molau2016results}
{Molau} S.,  {Crivello} S.,  {Goncalves} R.,  {Saraiva} C.,  {Stomeo} E.,
  {Kac} J.,  2016, WGN, Journal of the International Meteor Organization, \href
  {https://ui.adsabs.harvard.edu/abs/2016JIMO...44...92M} {44, 92}

\bibitem[\protect\citeauthoryear{{Molau}, {Crivello}, {Goncalves}, {Saraiva},
  {Stomeo}, {Strunk}  \& {Kac}}{{Molau} et~al.}{2018}]{molau2018results}
{Molau} S.,  {Crivello} S.,  {Goncalves} R.,  {Saraiva} C.,  {Stomeo} E.,
  {Strunk} J.,   {Kac} J.,  2018, WGN, Journal of the International Meteor
  Organization, \href {https://ui.adsabs.harvard.edu/abs/2018JIMO...46..205M}
  {46, 205}

\bibitem[\protect\citeauthoryear{Molau, Crivello, Goncalves, Saraiva, Stomeo,
  Strunk  \& Kac}{Molau et~al.}{2019}]{molau2019results}
Molau S.,  Crivello S.,  Goncalves R.,  Saraiva C.,  Stomeo E.,  Strunk J.,
  Kac J.,  2019, WGN, Journal of the International Meteor Organization, 47, 121

\bibitem[\protect\citeauthoryear{Moorhead, Egal, Brown, Moser  \&
  Cooke}{Moorhead et~al.}{2019}]{moorhead2019meteor}
Moorhead A.~V.,  Egal A.,  Brown P.~G.,  Moser D.~E.,   Cooke W.~J.,  2019,
  Journal of Spacecraft and Rockets, 56, 1531

\bibitem[\protect\citeauthoryear{Moorhead, Kingery  \& Ehlert}{Moorhead
  et~al.}{2020a}]{moorhead2020nasa}
Moorhead A.~V.,  Kingery A.,   Ehlert S.,  2020a, Journal of Spacecraft and
  Rockets, 57, 160

\bibitem[\protect\citeauthoryear{Moorhead, Clements  \& Vida}{Moorhead
  et~al.}{2020b}]{moorhead2020realistic}
Moorhead A.~V.,  Clements T.~D.,   Vida D.,  2020b, Monthly Notices of the
  Royal Astronomical Society, 494, 2982

\bibitem[\protect\citeauthoryear{Moorhead, Clements  \& Vida}{Moorhead
  et~al.}{2021}]{moorhead2021meteor}
Moorhead A.~V.,  Clements T.,   Vida D.,  2021, Monthly Notices of the Royal
  Astronomical Society, 508, 326

\bibitem[\protect\citeauthoryear{Murray, Hughes  \& Williams}{Murray
  et~al.}{1980}]{MurrayEtAl_QUA}
Murray C.~D.,  Hughes D.~W.,   Williams I.~P.,  1980, \mn@doi [Monthly Notices
  of the Royal Astronomical Society] {10.1093/mnras/190.4.733}, 190, 733

\bibitem[\protect\citeauthoryear{Musci, Weryk, Brown, Campbell-Brown  \&
  Wiegert}{Musci et~al.}{2012}]{musci2012optical}
Musci R.,  Weryk R.,  Brown P.,  Campbell-Brown M.,   Wiegert P.,  2012, The
  Astrophysical Journal, 745, 161

\bibitem[\protect\citeauthoryear{{Peterson}}{{Peterson}}{1999}]{Peterson_Spacecraft_Book}
{Peterson} G.,  1999, {Dynamics of Meteor Outbursts and Satellite Mitigation
  Strategies}.
The Aerospace Press

\bibitem[\protect\citeauthoryear{Pokorn\'{y} \& Brown}{Pokorn\'{y} \&
  Brown}{2016}]{pokorny2016reproducible}
Pokorn\'{y} P.,  Brown P.~G.,  2016, Astronomy \& Astrophysics, 592, A150

\bibitem[\protect\citeauthoryear{Poole, Hughes  \& Kaiser}{Poole
  et~al.}{1972}]{PooleEtAl_Quads}
Poole L. M.~G.,  Hughes D.~W.,   Kaiser T.~R.,  1972, \mn@doi [Monthly Notices
  of the Royal Astronomical Society] {10.1093/mnras/156.2.223}, 156, 223

\bibitem[\protect\citeauthoryear{Rendtel}{Rendtel}{2014}]{rendtel2014}
Rendtel J.,  2014, {Meteor Shower Workbook 2014}.
International Meteor Organization

\bibitem[\protect\citeauthoryear{{Rendtel}, {Koschack}  \& {Arlt}}{{Rendtel}
  et~al.}{1993}]{rendtel1993}
{Rendtel} J.,  {Koschack} R.,   {Arlt} R.,  1993, WGN, Journal of the
  International Meteor Organization, \href
  {https://ui.adsabs.harvard.edu/abs/1993JIMO...21...97R} {21, 97}

\bibitem[\protect\citeauthoryear{Rendtel, Ogawa  \& Sugimoto}{Rendtel
  et~al.}{2016}]{rendtel2016quadrantids}
Rendtel J.,  Ogawa H.,   Sugimoto H.,  2016, WGN, Journal of the International
  Meteor Organization, 44, 101

\bibitem[\protect\citeauthoryear{Rendtel et~al.,}{Rendtel
  et~al.}{2020}]{rendtel2020handbook}
Rendtel J.,  et~al., 2020, Handbook for meteor observers.
International Meteor Organization

\bibitem[\protect\citeauthoryear{Ryabova}{Ryabova}{2007}]{ryabova2007mathematical}
Ryabova G.,  2007, Monthly Notices of the Royal Astronomical Society, 375, 1371

\bibitem[\protect\citeauthoryear{Ryabova}{Ryabova}{2017}]{ryabova2017mass}
Ryabova G.,  2017, Planetary and Space Science, 143, 125

\bibitem[\protect\citeauthoryear{Ryabova}{Ryabova}{2021}]{Ryabova2021}
Ryabova G.~O.,  2021, \mn@doi [Monthly Notices of the Royal Astronomical
  Society] {10.1093/mnras/stab2286}, 507, 4481

\bibitem[\protect\citeauthoryear{Ryabova \& Rendtel}{Ryabova \&
  Rendtel}{2018}]{Ryabova2018}
Ryabova G.~O.,  Rendtel J.,  2018, \mn@doi [Monthly Notices of the Royal
  Astronomical Society: Letters] {10.1093/mnrasl/slx205}, 475, L77

\bibitem[\protect\citeauthoryear{Tabeshian, Wiegert, Ye, Hui, Gao  \&
  Tan}{Tabeshian et~al.}{2019}]{tabeshian2019asteroid}
Tabeshian M.,  Wiegert P.,  Ye Q.,  Hui M.-T.,  Gao X.,   Tan H.,  2019, The
  Astronomical Journal, 158, 30

\bibitem[\protect\citeauthoryear{Ulm}{Ulm}{1990}]{ulm1990simple}
Ulm K.,  1990, American journal of epidemiology, 131, 373

\bibitem[\protect\citeauthoryear{Vaubaillon, Rietze  \& Zilkova}{Vaubaillon
  et~al.}{2021}]{vaubaillon2021malbec}
Vaubaillon J.,  Rietze A.,   Zilkova D.,  2021, Monthly Notices of the Royal
  Astronomical Society, 508, 3897

\bibitem[\protect\citeauthoryear{{Verniani}}{{Verniani}}{1965}]{verniani1965}
{Verniani} F.,  1965, Smithsonian Contributions to Astrophysics, \href
  {https://ui.adsabs.harvard.edu/abs/1965SCoA....8..141V} {8, 141}

\bibitem[\protect\citeauthoryear{Verniani}{Verniani}{1973}]{verniani1973analysis}
Verniani F.,  1973, Journal of Geophysical Research, 78, 8429

\bibitem[\protect\citeauthoryear{Vida, Zubovi{\'c}, {\v{S}}egon, Gural  \&
  Cupec}{Vida et~al.}{2016}]{vida2016open}
Vida D.,  Zubovi{\'c} D.,  {\v{S}}egon D.,  Gural P.,   Cupec R.,  2016, in
  Proceedings of the International Meteor Conference (IMC2016), Egmond, The
  Netherlands. pp~2--5

\bibitem[\protect\citeauthoryear{Vida, Mazur, {\v{S}}egon, Zubovi{\'c},
  Kuki{\'c}, Parag  \& Macan}{Vida et~al.}{2018a}]{vida2018first}
Vida D.,  Mazur M.~J.,  {\v{S}}egon D.,  Zubovi{\'c} D.,  Kuki{\'c} P.,  Parag
  F.,   Macan A.,  2018a, WGN, Journal of the International Meteor
  Organization, 46, 2

\bibitem[\protect\citeauthoryear{Vida, Brown  \& Campbell-Brown}{Vida
  et~al.}{2018b}]{vida2018modelling}
Vida D.,  Brown P.~G.,   Campbell-Brown M.,  2018b, Monthly Notices of the
  Royal Astronomical Society, 479, 4307

\bibitem[\protect\citeauthoryear{Vida, Gural, Brown, Campbell-Brown  \&
  Wiegert}{Vida et~al.}{2020a}]{vida2020estimating}
Vida D.,  Gural P.~S.,  Brown P.~G.,  Campbell-Brown M.,   Wiegert P.,  2020a,
  Monthly Notices of the Royal Astronomical Society, 491, 2688

\bibitem[\protect\citeauthoryear{Vida, Campbell-Brown, Brown, Egal  \&
  Mazur}{Vida et~al.}{2020b}]{vida2020new}
Vida D.,  Campbell-Brown M.,  Brown P.~G.,  Egal A.,   Mazur M.~J.,  2020b,
  Astronomy \& Astrophysics, 635, A153

\bibitem[\protect\citeauthoryear{Vida et~al.,}{Vida
  et~al.}{2021}]{vida2021global}
Vida D.,  et~al., 2021, Monthly Notices of the Royal Astronomical Society, 506,
  5046

\bibitem[\protect\citeauthoryear{Voj{\'a}{\v{c}}ek, Borovi{\v{c}}ka, Koten,
  Spurn{\`y}  \& {\v{S}}tork}{Voj{\'a}{\v{c}}ek
  et~al.}{2019}]{vojavcek2019properties}
Voj{\'a}{\v{c}}ek V.,  Borovi{\v{c}}ka J.,  Koten P.,  Spurn{\`y} P.,
  {\v{S}}tork R.,  2019, Astronomy \& Astrophysics, 621, A68

\bibitem[\protect\citeauthoryear{Wiegert \& Brown}{Wiegert \&
  Brown}{2005}]{wiegert2005}
Wiegert P.,  Brown P.,  2005, Icarus, 179, 139

\bibitem[\protect\citeauthoryear{Zvolankova}{Zvolankova}{1983}]{zvolankova1983dependence}
Zvolankova J.,  1983, Bulletin of the Astronomical Institutes of
  Czechoslovakia, 34, 122

\bibitem[\protect\citeauthoryear{de Le{\'o}n, Campins, Tsiganis, Morbidelli  \&
  Licandro}{de~Le{\'o}n et~al.}{2010}]{de2010origin}
de Le{\'o}n J.,  Campins H.,  Tsiganis K.,  Morbidelli A.,   Licandro J.,
  2010, Astronomy \& Astrophysics, 513, A26

\makeatother
\end{thebibliography}

\bsp	
\label{lastpage}
\end{document}